\documentclass[a4paper]{article}
\pdfoutput=1
\usepackage[round]{natbib}
\usepackage{amsmath,amsthm,amssymb}
\usepackage{float}
\usepackage{physics}
\usepackage{graphicx}
\usepackage{setspace}
\usepackage{subcaption}
\usepackage{fullpage}
\usepackage{authblk}
\usepackage{lineno}
\usepackage{enumitem}
\usepackage{hyperref}
\hypersetup{
    colorlinks=true,
    citecolor=blue,
}

\begin{document}

\title{A Novel CFD-DEM Coarse-Graining Method Based on the Voronoi Tessellation}
\author[1]{Hanqiao Che\footnote{Corresponding author, email: \href{h.che@imperial.ac.uk}{h.che@imperial.ac.uk}}}
\author[1]{Catherine O'Sullivan}
\author[2]{Adnan Sufian}
\author[3]{Edward Smith}
\affil[1]{Department of Civil and Environmental Engineering, Imperial College London, South Kensington Campus, SW7 2AZ, United Kingdom}
\affil[2]{School of Civil Engineering, The University of Queensland, St Lucia, 4072, Australia}
\affil[3]{Department of Mechanical and Aerospace Engineering, Brunel University London, Uxbridge, Middlesex UB8 3PH, United Kingdom}
\date{}
\setcounter{Maxaffil}{0}
\renewcommand\Affilfont{\itshape\small}

\maketitle

\providecommand{\keywords}[1]
{
  \small    
  \textbf{\textit{Keywords---}} #1
}

\begin{abstract} 
\noindent In unresolved flow CFD-DEM simulations, the porosity values for each CFD cell are determined using a coarse-graining algorithm. While this approach enables coupled simulations of representative numbers of particles, the influence of the porosity local to the particles on the fluid-particle interaction force is not captured. This contribution considers a two-grid coarse-graining method that determines a local porosity for each particle using a radical Voronoi tessellation of the system. A relatively fine, regular point cloud is used to map these local porosity data to the CFD cells. The method is evaluated using two different cases considering both disperse and tightly packed particles. The data show that the method conserves porosity data, is reasonably grid-independent and can generate a relatively smooth porosity field. The new method can more accurately predict the fluid-particle interactive force for polydisperse particle system than alternative methods that have been implemented in unresolved CFD-DEM codes.

\vspace{5mm}

\noindent \keywords{Radical Voronoi tessellation, CFD-DEM, Coarse-Graining, Fluid-particle interactive force}
\end{abstract}

\setlength{\parindent}{0em}
\setlength{\parskip}{1em}

\section{\label{sec:1}Introduction}

Coupled computational fluid dynamics - discrete element method (CFD-DEM) simulations can provide valuable insight across a range of applications including the granulation of pharmaceutical products \citep{heinrich2015}, mining/mineral extraction \citep{han2003}, debris flows and internal erosion of dams and flood embankments \citep{hu2019}. 
A common approach to CFD-DEM is based on an Eulerian-Lagrangian framework, where the fluid phase is simulated using a fixed Eulerian grid with grid spacing that is larger than the particles (unresolved approach) or measurably smaller than the particles (resolved approach) \citep{zhou2010}. 
One resolved CFD-DEM approach is the immersed boundary method (IBM) \citep{peskin2002} in which the points on the surfaces of particles are inserted in the fluid region as additional no-slip boundary conditions, and a source term representing the fluid-particle interaction is added into the Navier-Stokes equations to describe the fluid flow that is solved on the fixed Eulerian grid. 
In unresolved CFD-DEM, a porous version of the Navier-Stokes equations is solved \citep{anderson1967}; the presence of particles in the fluid is accounted for by including the porosity value. 
The particle-fluid interaction forces are estimated by empirical drag correlations \citep{ergun1952,gidaspow1994,tenneti2011,tang2015} and included in the CFD equations as a source term. 
On the DEM side the particle-fluid interaction force is considered alongside the inter-particle contact forces and any gravitational forces \citep{zhou2010,kloss2012} when determining the resultant forces acting on the particles. 
CFD-DEM is a coupled simulation process and the DEM and CFD solvers can be executed either concurrently or sequentially. 
After a specified number of time-steps, relevant variables such as porosity, drag force and velocity of each phase are exchanged between both solvers.

In an unresolved CFD-DEM simulation, the coarse-graining stage is very important. In this stage, the relevant coupling data, including the porosity, solid velocity and the momentum source terms for each CFD cell, are determined. 
The coarse-graining approach used must satisfy several requirements \citep{sun2015,clarke2018}, most significantly it must (1) conserve relevant physical quantities such as the total particle volume; (2) be able to achieve relatively grid-independent results, and (3) be able to produce smooth coarse-grained fields without large jumps existing between the neighbouring CFD cells. 

The simplest coarse-graining method adopted in unresolved CFD-DEM is the particle centroid method (PCM) \citep{tsuji1993}. 
PCM estimates the volume of the particles within a CFD cell by summing up the volume of the particles whose centres are located in that cell. 
This method is fast but can be only applied in cases where the cell-to-particle size ratio is large. 
Large jumps in the porosity values may occur and lead to big oscillations and high-frequency peaks in the simulation results \citep{peng2014}. 

In the divided method (DM) \citep{kloss2012,wu2009} the proportion of each particle is assigned to the CFD cell is based on the actual overlap volume between the particle and the cell boundary; this is more accurate than the PCM approach. 
There are many strategies to estimate the overlap volume. 
For example, the particle can be first resolved by a series of distributed marker points to evenly apportion the particle’s volume to all cells that are (partly) covered, and then the relevant particle sub-region is assigned to the CFD cell using PCM \citep{kloss2012,peng2016}. 
Where the particle size is approximately equal to or even larger than the grid size, the large particles can be represented by a number of small particles \citep{tsuji2014}. 
DM can generally eliminate the jumps in the porosity field between the neighbouring CFD cells and provides smoother porosity field than PCM.

The porous sphere \citep{jing2016} or porous cube \citep{link2005} methods represent the particles as porous elements and then distribute the volume of the elements to the CFD cells within a region that is larger than the particle size to avoid extreme porosity values in the CFD cells. 
To further improve the smoothness of the porosity field, the statistical kernel method \citep{glasser2001} or the diffusion-based coarse-graining method \citep{sun2015} can be used. 
These methods apply a statistical kernel or a diffusion source to the centre of each particle, and the volume of the particle is distributed to the calculation domain. 
In these approaches, a portion of the particle’s volume can be assigned to a fluid cell, even though the particle does not intersect or overlap with that cell. 
Therefore, a smooth porosity field can be obtained even when the CFD cell size is smaller than the particle diameter \citep{sun2015}. 

The general approach adopted in the above-mentioned methods is to firstly bin or assign a proportion of each particle’s volume to the predefined fluid cells, and then calculate the porosity of each fluid cell. 
Accordingly, the porosity data are “grid-dependent”. The CFD grid employed should be chosen to maintain the smoothness of the porosity field and the resolution of fine-scale fluid features is limited \citep{deb2013}. 
A two-grid method has been proposed by \citet{deb2013} and \citet{su2015}, in which the porosity is calculated based on a coarse grid (grid size larger than particles) and mapped to a finer grid that is used by the CFD solver. 
However, a key issue with this type of two-grid method is how to select an appropriate grid size. 
The shape and size of the coarse grid selected seems arbitrary, for example, a cube with three times the particle diameter as the length was used by \citet{deb2013} and no data to support decisions around the choice of grid spacing is available.

\begin{figure}[ht]
    \centering
    \begin{subfigure}[b]{0.49\textwidth}
        \centering
        \includegraphics[width=5cm]{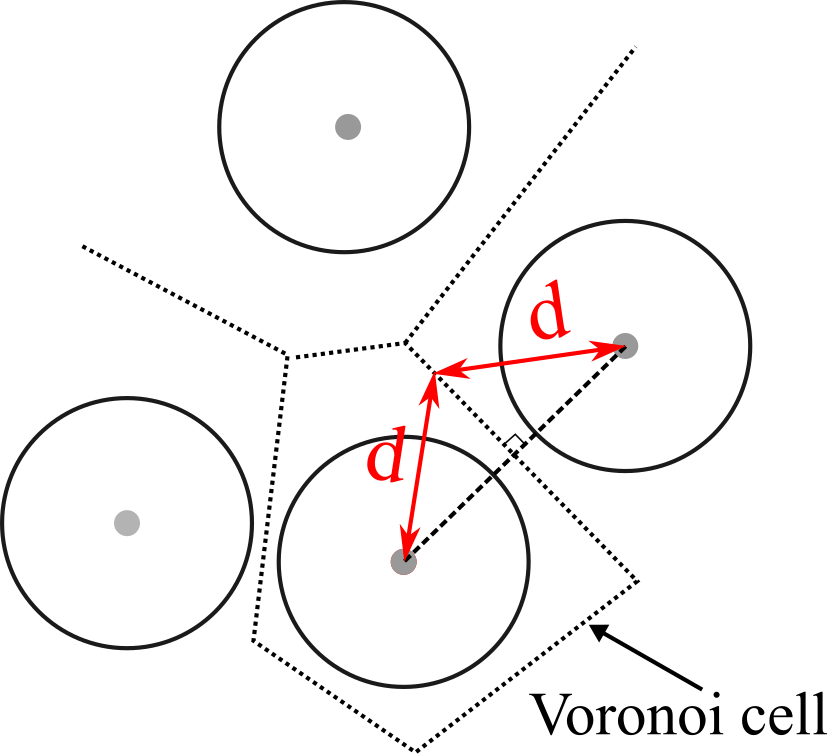}
        \caption{\label{fig:1a}Conventional Voronoi tessellation}
    \end{subfigure}
    \hfill
    \begin{subfigure}[b]{0.49\textwidth}
        \centering
        \includegraphics[width=4.5cm]{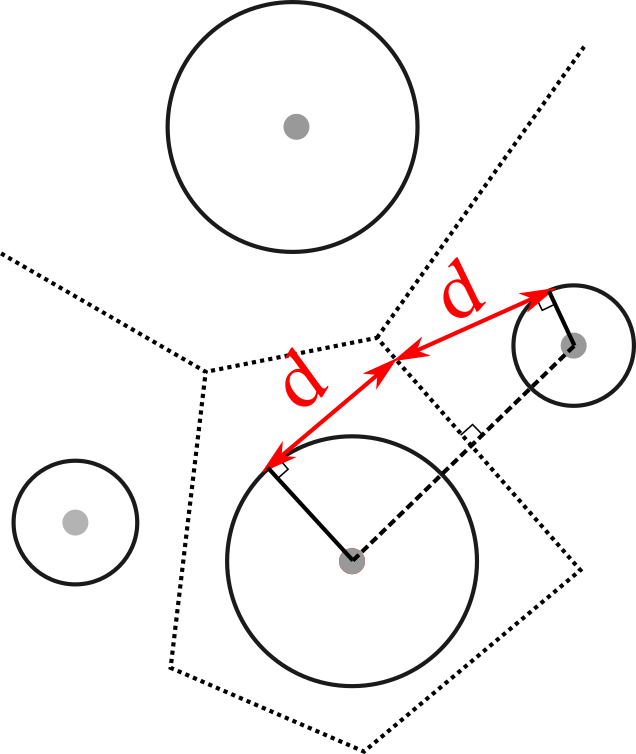}
        \caption{\label{fig:1b}Radical Voronoi tessellation}
    \end{subfigure}
    \caption{\label{fig:1}Schematic of the conventional and radical Voronoi tessellation}
\end{figure}

A Voronoi tessellation can be used to identify a local porosity for each particle in the system. 
A Voronoi tessellation divides the particle domain into a unique space filling set of polyhedral cells (Fig.~\ref{fig:1a}). 
In case of monodisperse particle systems, only the coordinates of the particle centre are used as the input; the planes of the polyhedral Voronoi cells are located at the midpoints of the lines (Delaunay tetrahedra edges) connecting adjacent particles. 
For polydisperse particle systems, if this basic partitioning approach is adopted the polyhedral planes defining the Voronoi cells can intercept the particles. 
The radical Voronoi tessellation was developed to solve this problem \citep{gellatly1982}. 
In a radical Voronoi tessellation, the space between particles is partitioned based on radical planes, i.e., the planes formed between pairs of particles for which the tangent lines from a point in the plane to both particles are of equal length. 
Voronoi tessellations can be implemented in DEM codes using existing libraries \citep{rycroft2009}. 
This approach has been used in previous DEM simulations to quantify the void space \citep{frenning2015}. 
By comparing with IBM simulations, \citet{knight2020} showed that, in principle, this local porosity could be used to calculate the drag force on individual particles in polydisperse systems. 
However, there are no published studies documenting a successful implementation of this approach in an unresolved CFD-DEM simulation framework.

The objective of this paper is to introduce a new CFD-DEM coarse-graining method and evaluate its performance. 
Following an overview of unresolved CFD-DEM, Section~\ref{sec:3} describes how the local porosity of each particle is calculated based on the particle volume and the volume of the surrounding Voronoi cells, and then mapped to the CFD grid using a point cloud method. 
In Section~\ref{sec:4} two validation cases which represent typical applications of CFD-DEM are considered, and the simulation data generated are compared to previously generated IBM data from \citet{knight2020} as well as to appropriate experimental data.

\section{\label{sec:2}Governing equations in CFD-DEM}

\begin{figure}[ht]
    \centering
    \includegraphics[width=15cm]{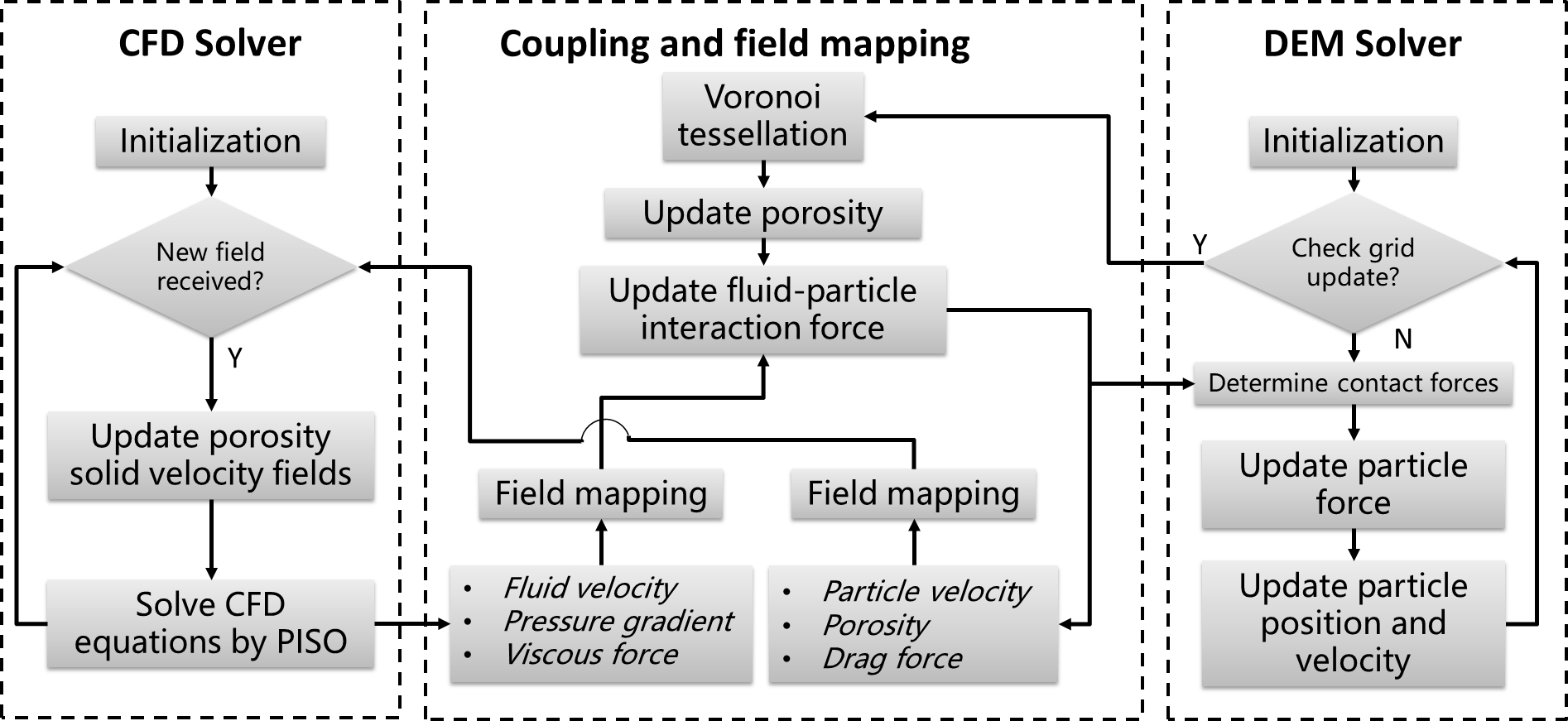}
    \caption{\label{fig:2}CFD-DEM coupling in the CPL-Library}
\end{figure}

CFD-DEM typically involves coupling an established DEM code with an established CFD solver \citep{sun2016,goniva2010,kloss2009}. 
In the current study, the open-source code CPL-Library \citep{smith2020} was used to couple the open-source codes LAMMPS and OpenFOAM. 
All coupling in CPL-Library is handled through shared libraries which use the message passing interface (MPI) to facilitate information exchange between software. 
Fig.~\ref{fig:2} shows the flowchart of the CFD-DEM coupling in CPL-Library; the Navier-Stokes equations are solved in the CFD solver, while the DEM solver simulates the motion of and interaction between the particles. 

In unresolved CFD-DEM simulations, the porosity ($\varepsilon_f$) is included in the system of equations considered in the CFD solver. 
For incompressible flow, the volume-averaged continuity and momentum equations are written as:
\begin{equation}
    \frac{\partial \varepsilon_f}{\partial t} + \nabla \cdot \left( \varepsilon_f \mathbf{u}_f \right) = 0
\end{equation}
\begin{equation}
    \frac{\partial}{\partial t}\left( \varepsilon_f \rho_f \mathbf{u}_f \right)
    +
    \nabla\cdot\left( \varepsilon_f \rho_f \mathbf{u}_f \otimes \mathbf{u}_f \right)
    = 
    -
    \varepsilon_f \nabla p
    +
    \varepsilon_f \nabla \cdot \mathbf{\tau}
    -
    \mathbf{S}_{pf}
    +
    \varepsilon_f \rho_f \mathbf{g}
\end{equation}
where $\rho_f$ and $\mathbf{u}_f$ are the fluid density and velocity, $\mathbf{\tau}$ is the viscous stress tensor, $\mathbf{g}$ is gravity. 
$\mathbf{S}_{pf}$ is the momentum source term resulted from the fluid-solid interaction; for reasons of numerical stability, it is split-up into an implicit and an explicit part \citep{kloss2012,sun2016} and is written as
\begin{equation}
    \label{eq:3}
    \mathbf{S}_{pf} = K_{pf}\mathbf{u}_f - K_{pf}\left<\mathbf{u}_p\right>
\end{equation}
where $\left<\mathbf{u}_p\right>$ is the cell-based ensemble-averaged particle velocity, $\mathbf{u}_f$ is discretised implicitly in the CFD solver; the (scalar) coefficient $K_{pf}$ is a function of the drag force on particles, which is estimated by
\begin{equation}
    K_{pf} = \frac{\abs{\sum_i \mathbf{F}_{d,i}}}{V_{cell}\abs{\mathbf{u}_f - \left<\mathbf{u}_p\right>}}
\end{equation}
where $\mathbf{F}_{d,i}$ is the drag force acting on a single particle $i$. 
There are many empirical drag correlations available to estimate $\mathbf{F}_{d,i}$. 
In the validation examples below the \citet{ergun1952}, \citet{tenneti2011} and \citet{wen1966} drag models are considered. 
The PISO (Pressure Implicit with Split Operator) pressure–velocity coupling algorithm \citep{issa1986} is adopted to solve the above continuity and momentum equations.

In the DEM solver the motion of particle $i$ is calculated by the total force and torque acting on it:
\begin{equation}
    m_i \frac{\mathrm{d} \mathbf{u}_i}{\mathrm{d}t}
    =
    m_i \mathbf{g}
    +
    \sum_{j=0}^k \left(F_{n,ij}\mathbf{n} + F_{t,ij}\mathbf{t} \right)
    +
    \mathbf{F}_{d,i}
    +
    \mathbf{F}_{p,i}
    +
    \mathbf{F}_{\tau,i}
\end{equation}
\begin{equation}
    \mathbf{I}_i \frac{\mathrm{d} \omega_i}{\mathrm{d}t}
    =
    \sum_{j=0}^k \left[\mathbf{r} \times \left(F_{t,ij}\mathbf{t}\right) \right]
\end{equation}
where $F_{n,ij}$ and $F_{t,ij}$ are the normal and tangential contact forces associated with contact between particles $i$ and $j$, $k$ is the total number of contacts involving particle $i$, $\mathbf{F}_{p,i}$ and $\mathbf{F}_{\tau,i}$ represent the pressure gradient force and the viscous force respectively, which are exerted by the surrounding fluid; $\mathbf{I}_i$ is the moment of inertia of the particle. 
The magnitudes of the contact force are calculated based on a spring-dashpot soft sphere contact model with a “history” effect \citep{cundall1979}:
\begin{equation}
    F_n = -k_n \delta_n + c_n \Delta u_n
\end{equation}
\begin{equation}
    F_t
    =
    \begin{cases} 
    k_t \int_{t_0}^t \Delta u_t \mathrm{d} t + c_t \Delta u_t &\mathrm{for\:} \abs{F_t} < \mu F_n \\
    \mu F_n &\mathrm{for\:} \abs{F_t} \geq \mu F_n
    \end{cases}
\end{equation}
where $k_n$ and $k_t$ are the normal and tangential spring stiffnesses, $c_n$ is the normal viscous dashpot coefficient and $\Delta u_t$ is the incremental tangential displacement. 
Here $k_n$ and $k_t$ are determined using the simplified Hertz-Mindlin contact model \citep{che2020}.

The coupling loop proposed here includes 4 main steps: 
\emph{Step 1}: The DEM and CFD solver are executed concurrently; 
\emph{Step 2}: A radical Voronoi tessellation is carried out based on the current particle positions so that the local particle porosity can be evaluated; 
\emph{Step 3}: The fluid fields are mapped to each particle and the drag force acting on each particle is calculated; 
\emph{Step 4}: The particle porosity, velocity and the drag force are mapped to the CFD side; 
\emph{Step 5}: Based on the updated fluid information for coupling, the routine is repeated from Step 1. 
The tessellation used in Step 2 and the mappings used in Steps 3 and 4 are described in the following sections.

\section{\label{sec:3}Two-grid, Voronoi tessellation-based coarse-graining method}

The two-grid coarse-graining method proposed here comprises two main steps: 
(1) Radical Voronoi tessellation of the particle system to assign a local porosity to each particle; 
(2) Field mapping to the CFD grid using a point cloud.

\subsection{\label{sub:3.1}Voronoi tessellation of the particle system}

\begin{figure}[ht]
    \centering
    \begin{subfigure}[b]{0.49\textwidth}
        \centering
        \includegraphics[width=4.75cm]{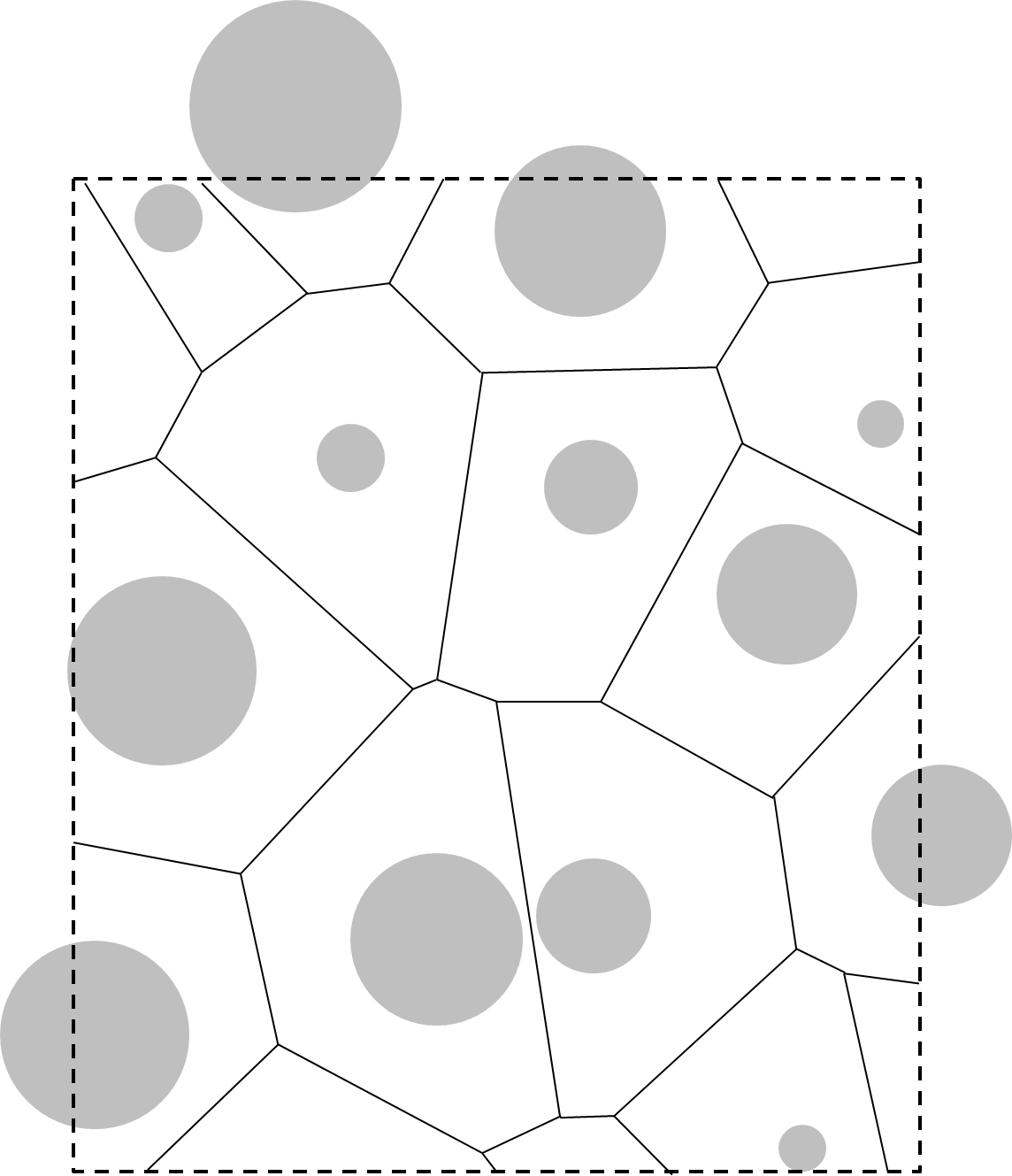}
        \caption{\label{fig:3a}Uniform particle packing}
    \end{subfigure}
    \hfill
    \begin{subfigure}[b]{0.49\textwidth}
        \centering
        \includegraphics[width=4.75cm]{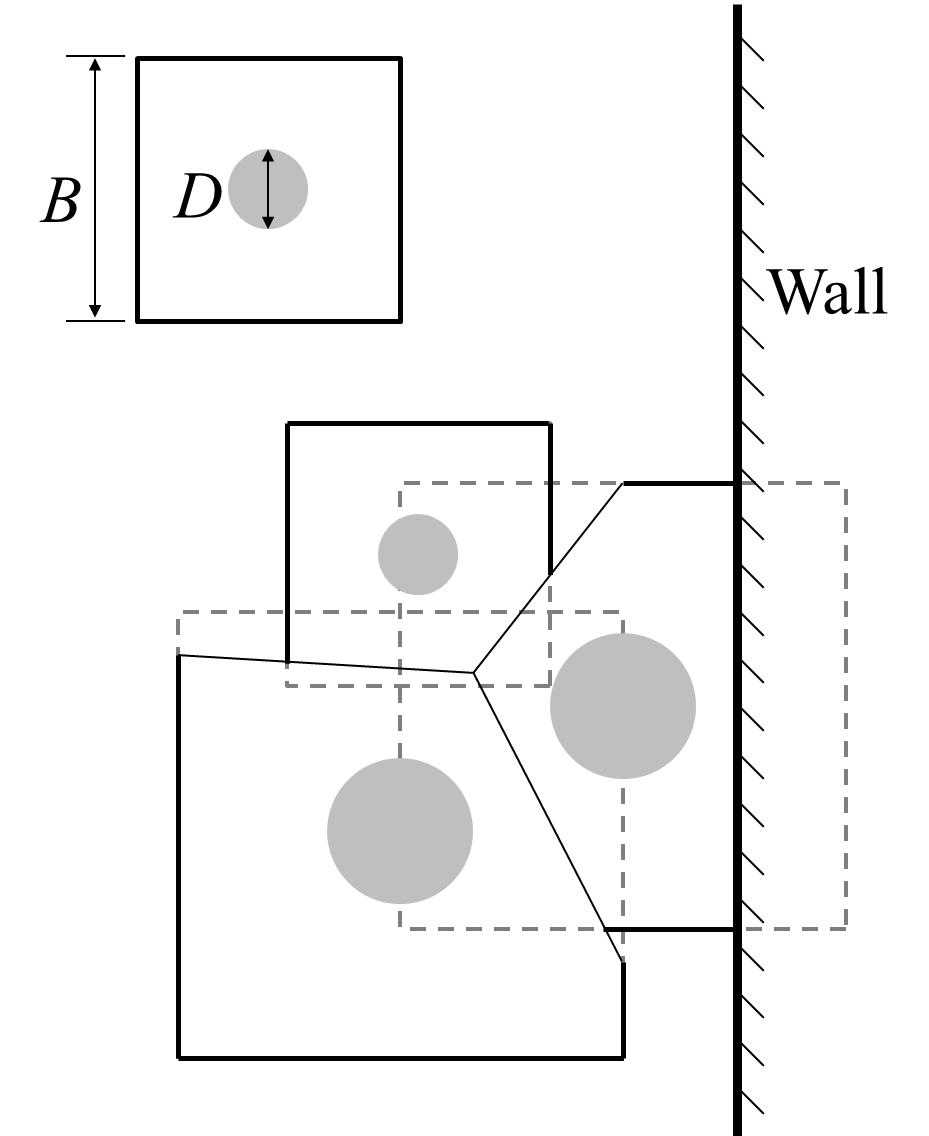}
        \caption{\label{fig:3b}Dilute particle flow region}
    \end{subfigure}
    \caption{\label{fig:3}Schematic of the Voronoi tessellation of the particle system (the Voronoi cells are built in 3D space)}
\end{figure}

Fig.~\ref{fig:3} is a schematic of the Voronoi tessellation of the particle system. 
In the case of a uniformly distributed particle system in a predefined domain, use of a Voronoi tessellation to obtain representative local porosity values is straightforward (see Fig.~\ref{fig:3a}). 
However, in many applications, such as fluidised beds, there are regions with dilute particle flow. In these regions the Voronoi cells become large and would give unreasonable estimates of the local porosity. 
To resolve this problem, a bounding cuboid which is larger than the particle is generated. 
If the boundaries of the particle’s Voronoi cell stretches beyond the initial bounding cuboid, they are replaced by the cuboid boundaries (see Fig.~\ref{fig:3b}).
The size of the bounding cuboid is controlled by a parameter ($\theta_1$), which equals the cuboid-to-particle size ratio ($\theta_1 = B/D$). 
The cuboid bounding box gives an upper bound to the calculated porosity, for example, for $\theta_1$ = 2, 3 and 4, the maximum local porosities for individual particles are 0.93, 0.98 and 0.99, respectively.

The local porosity of an individual particle is simply the volume of the void space within the particle’s Voronoi cell divided by the total volume of the particle’s Voronoi cell, i.e.,
\begin{equation}
    \varepsilon_{f,i}^V = \frac{V_{voro,i} - V_{p,i}}{V_{voro,i}}
\end{equation}
where $V_{p,i}$ and $V_{voro,i}$ are the volumes of particle $i$ and the surrounding Voronoi cell, the superscript $V$ represents the porosity is in a Voronoi cell.

Here the Voronoi tessellation of the particle system was carried out by adding \texttt{voro++} library to LAMMPS as a user package \citep{rycroft2009}, and it can be compiled with the CPL-Library for the CFD-DEM coupling.

\subsection{\label{sub:3.2}Field mapping using the point cloud}

\begin{figure}[ht]
    \centering
    \begin{subfigure}[b]{0.49\textwidth}
        \centering
        \includegraphics[width=8cm]{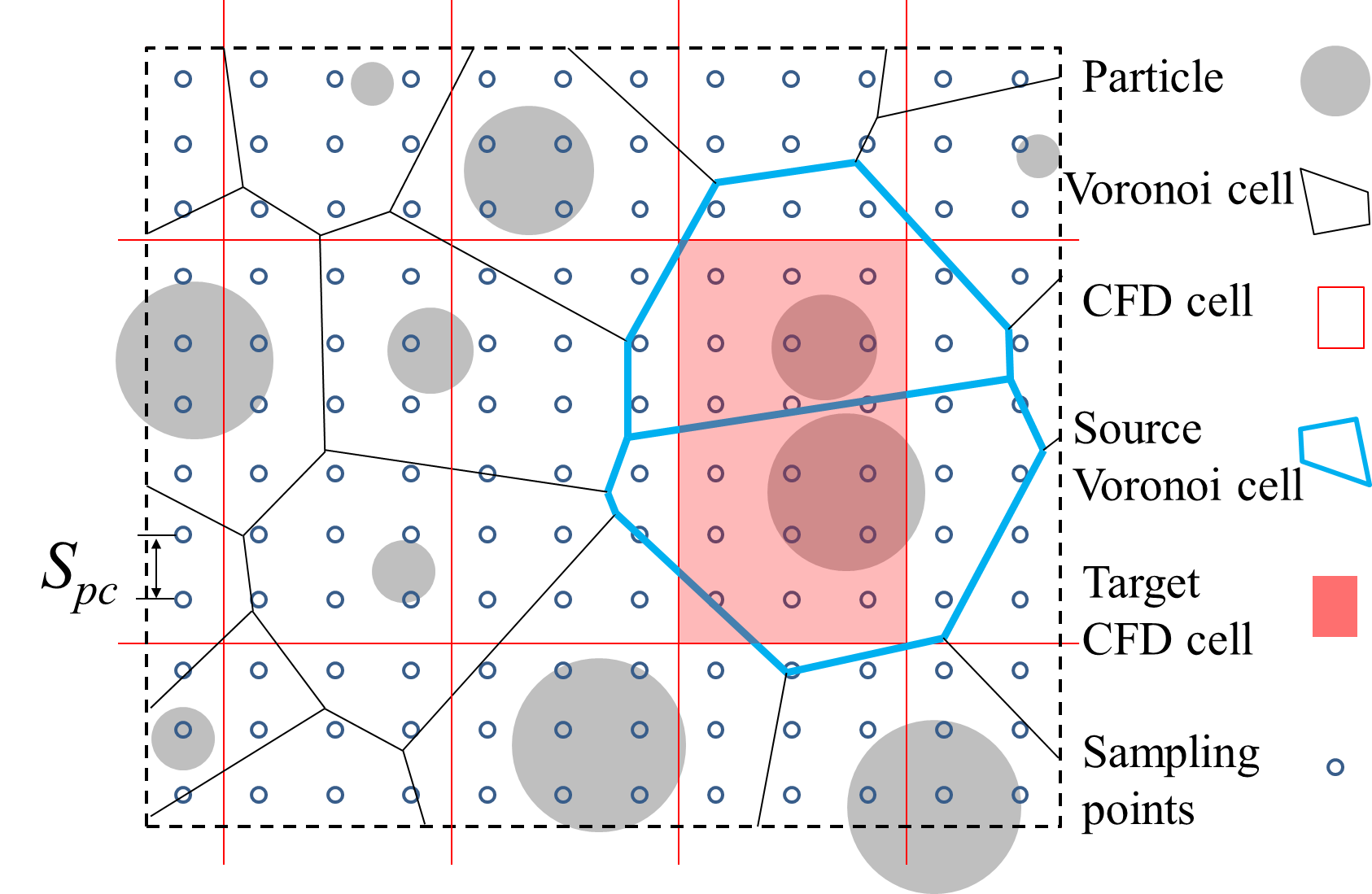}
        \caption{\label{fig:4a}Field mapping from Voronoi cells to a CFD cell}
    \end{subfigure}
    \hfill
    \begin{subfigure}[b]{0.49\textwidth}
        \centering
        \includegraphics[width=7.5cm]{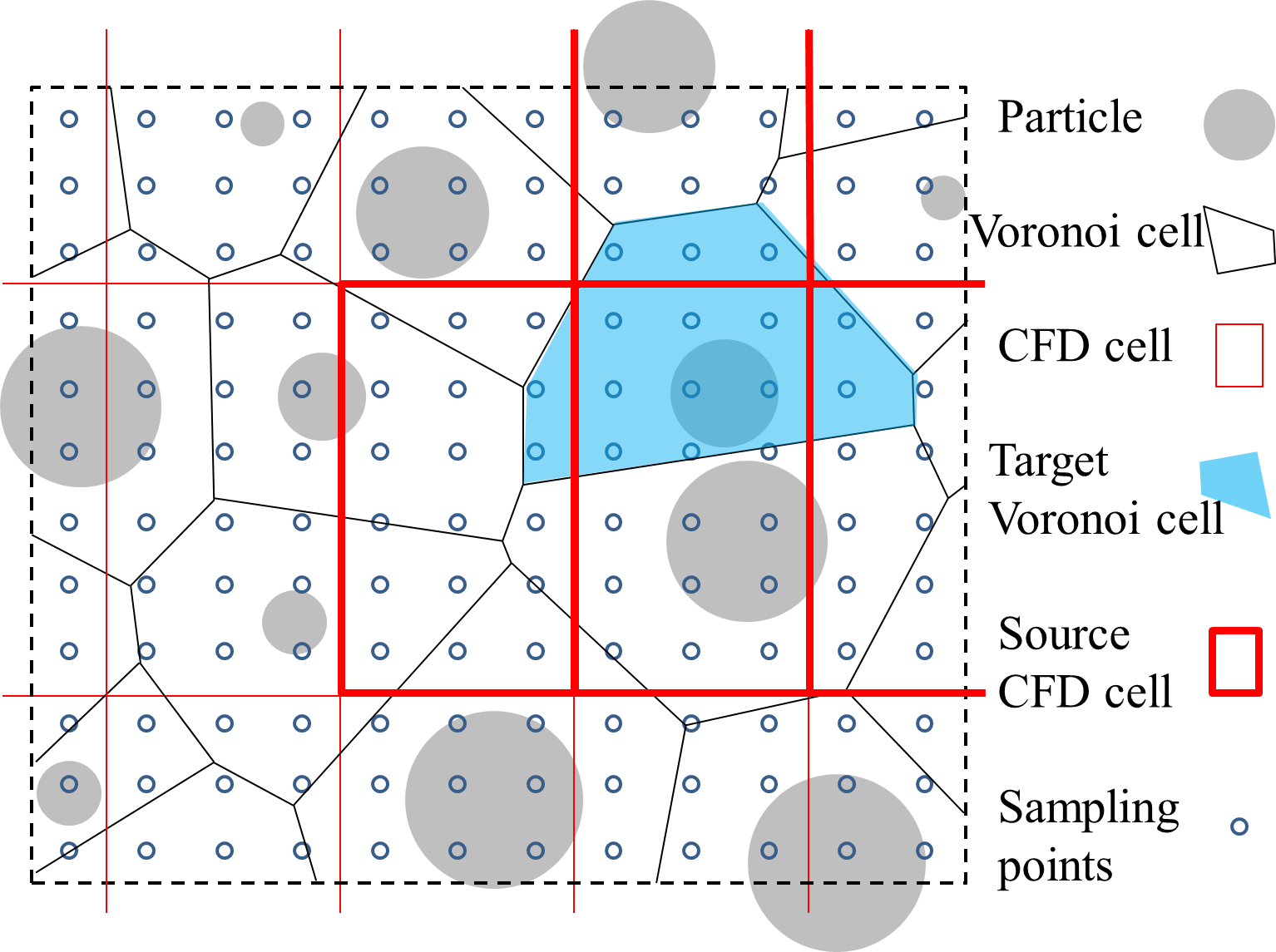}
        \caption{\label{fig:4b}Field mapping from CFD cells to a Voronoi cell}
    \end{subfigure}
    \caption{\label{fig:4}Schematic of the field mapping method using point cloud.}
\end{figure}

As noted above, CFD-DEM requires a field mapping to exchange key information between the DEM and CFD solvers. 
Due to the complex geometric structure of the Voronoi cells, the field mapping of the particle data to the fluid grid is achieved by mapping data to a point cloud. 
This method uses a fine uniform grid of discrete points (termed a point cloud) to identify (with some approximation/discretization error) the region that is common to a particular Voronoi cell and the CFD cell that overlaps it. 
As shown in Fig.~\ref{fig:4}, a regular grid of sampling points is generated that is evenly distributed across the whole computational domain. 
The density of the point cloud is denoted by the parameter $\theta_2$, which equals the ratio of the smallest particle diameter and the point cloud grid spacing ($\theta_2 = D_{min}/S_{pc}$). 
A systematic inspection of the sampling points in the cloud (with given position vectors) is carried out to identify which particle Voronoi cell and which CFD cell contain the point. 
This exercise generates two one-dimensional arrays (vectors); these store (1) the number of sampling points lying within Voronoi cell $i$ ($N^{total}_i$) and (2) the number of sampling points in CFD cell $n$ ($N^{total}_n$). 
Then a 2D sparse array that stores the number of sampling points shared by specific Voronoi cell $i$ and CFD cell $n$ ($N^{share}_{i,n}$) is populated. 
The arrays $N^{total}_i$, $N^{total}_n$ and $N^{share}_{i,n}$ are updated whenever the Voronoi cells are rebuilt and are used in the subsequent field mapping procedures.

Fig.~\ref{fig:4a} shows the schematic of the forward mapping step, i.e., the mapping from a “source” Voronoi cell to the overlying “target” CFD cell. 
For a given porosity field $\varepsilon^V_{f,i}$ in the Voronoi polygon $i$, the magnitude of the porosity mapped to target CFD cell $n$ is given by
\begin{equation}
    \varepsilon_{f,n}^C = \sum_{i=1}^{N_p} \left( \varepsilon^V_{f,i} \frac{N^{share}_{i,n}}{N^{total}_n} \right)
\end{equation}
where $N_p$ equals the total number of particles in the simulation domain and the superscript $C$ indicates that the calculated porosity relates to the CFD grid. 
For the particle velocity, the mapping scheme is
\begin{equation}
    u_{s,n}^C 
    =
    \frac
    {\sum_{i=1}^{N_p} \left( 1 - \varepsilon^V_{f,i} \right) u_{s,i}^V N^{share}_{i,n}}
    {\sum_{i=1}^{N_p} \left( 1 - \varepsilon^V_{f,i} \right) N^{share}_{i,n}} 
\end{equation}
where $u_{s,i}^V$ and $u_{s,n}^C$ are the particle velocities in Voronoi cell $i$ and CFD cell $n$, respectively. 
If $\mathbf{K}_n$ represents either the implicit or explicit part of the momentum source term in a CFD cell (Equation~\ref{eq:3}), and $\mathbf{F}_i$ represents the corresponding part of the drag force acting on particle $i$, then the forward mapping of the drag forces from the DEM solver to the CFD solver is
\begin{equation}
    \mathbf{K}_n 
    =
    \sum_{i=1}^{N_p}
    \frac
    {N^{share}_{i,n} \mathbf{F}_i}
    {N^{total}_n V_{voro,i}} 
\end{equation}
The schematic of the backward mapping (from CFD to DEM) is shown in Fig.~\ref{fig:4b}. 
For the field of $\chi^C_n$, which can be the fluid velocity, pressure gradient force or viscous force, in CFD cell $n$, the mapping scheme adopts a similar logic as the forward step so that
\begin{equation}
    \chi_i^V 
    =
    \sum_{n=1}^{N_f}
    \chi_n^C
    \frac
    {N^{share}_{i,n}}
    {N^{total}_i} 
\end{equation}
where $N_f$ is the total number of the CFD grid cells and $\chi^V_i$ is the field variable mapped to Voronoi polygon $i$.

To improve the calculation efficiency in both the forward and backward mapping, the field mapping procedure loops over the cells in a linked list (a “neighbour list”), which records the cells in the other grid that have a non-zero $N^{share}_{i,n}$ with the target cell (i.e. for each fluid cell there is a neighbour list of Voronoi cells). 
In this way, less than 1\% of the total number of cells in the other grid are need to be considered for a given target cell.

In the current version of CPL-Library, only uniform (regular) block meshes are available in the CFD solver (see Fig.~\ref{fig:4}), and so the field mapping procedures proposed here were implemented and tested for the case of uniform (regular) meshes. 
However, the generalization of the method proposed here to a non-uniform CFD mesh topology is straight-forward.

The distance between the neighbouring sampling points should be smaller than the edge length of a hexahedron inscribed in the smallest particle ($\theta_2 < 0.577 D_{min}$), so that each Voronoi cell contains at least one sampling point. 
The effect of the point cloud density on the field mapping is considered below.

There are two main advantages to this approach. 
Firstly, the volume of the Voronoi cell is always larger than the particle, so that all of the CFD and Voronoi cells will have non-zero porosity. 
Secondly, as the field mapping procedure generates volume-weighted field values using data from several adjacent cells in the other grid, a smooth field can be obtained, which is also a merit of the conventional two-grid methods. 

\subsection{\label{sub:3.3}Simulation data assessment method}

To quantitatively assess the porosity field obtained from different coarse-graining methods, three measures are considered. The first is the relative error ($Er$) of the total particle volume that is mapped onto the CFD grid, $V^C_{p,total}$, to the real solid volume, $V_{p,total}$,  (i.e. the total volume of particles in the DEM solver), 
\begin{equation}
    Er =\frac{V^C_{p,total} - V_{p,total}}{V_{p,total}} 
\end{equation}
The second measure is the standard deviation ($SD$) of the porosity distributions which is defined as
\begin{equation}
    SD = \sqrt{\frac{1}{N_{occ}-1} \sum_{n=1}^{N_{occ}} \left( \varepsilon_{f,n}^C -\overline{\varepsilon_f^C} \right)^2} 
\end{equation}
where $N_{occ}$ is the total number of CFD cells that are occupied by the par{}ticles. 
Following \citet{xie1994} and \citet{yang2002}, the third measure considered is the residual image error ($\gamma$) which compares the difference between two similar images in terms of pixel intensity level and is given by
\begin{equation}
    \gamma =\frac{\abs{\boldsymbol{\hat\varepsilon}_f - \boldsymbol{\varepsilon}_f}}{\abs{\boldsymbol{\varepsilon}_f}} 
\end{equation}
where $\boldsymbol{\hat\varepsilon}_f$ and $\boldsymbol{\varepsilon}_f$ are the calculated and the reference image vectors of the porosity, respectively. 
In the following section, the image error considers $\gamma$ for neighbouring frames in a sequence of porosity images obtained from the calculation.

The Pearson correlation coefficient ($\rho$) was introduced to assess the relationship between the individual particle drag forces obtained from different calculation methods. 
The Pearson correlation coefficient between two variables X and Y is calculated by \citep{zaiontz2002}
\begin{equation}
    \rho_{X,Y} = \frac{\mathrm{E}\left[ \left(X - \overline{X} \right) \left(Y - \overline{Y} \right) \right]}{SD_X SD_Y}
\end{equation}
where $SD_X$ and $SD_Y$ are the standard deviation of $X$ and $Y$, $\overline{X}$ and $\overline{Y}$ are mean values, and $E$ represents the expectation.

\section{\label{sec:4}Results and discussion}

Two typical but contrasting cases were selected to assess the performance of the proposed method. 
Case 1 is a uniformly packed polydisperse particle assembly with periodic boundaries, this type of assembly can be used to evaluate the fluid-particle interactions in the field of soil mechanics \citep{hu2019,huang2014}.
Case 2 is a spouted fluidised bed, which involves the flow of monodisperse particles with distinctive solid concentration distributions. 
This type of fluidised bed is frequently seen in the pharmaceutical industry to coat particles \citep{defreitas2019}. 
The new method proposed here is denoted 2GVM (Two-grid Voronoi Method) and the calculation results are compared to those obtained using PCM and DM \citep{link2005}.

\subsection{\label{sub:4.1}Validation Case 1: flow through particle assembly}

\begin{figure}[b]
    \centering
    \begin{subfigure}[b]{0.49\textwidth}
        \centering
        \includegraphics[width=7cm]{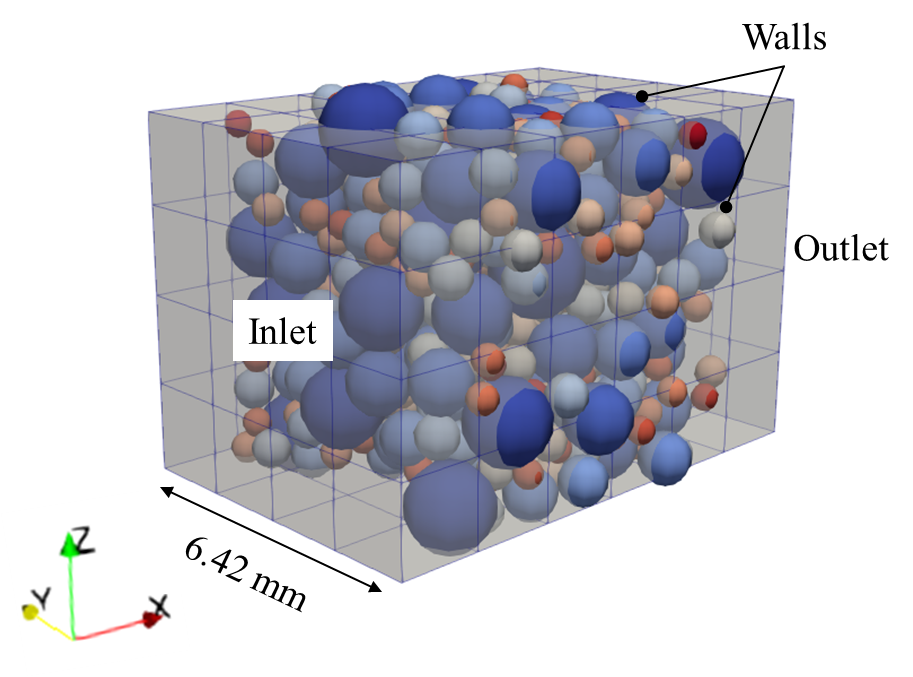}
        \caption{\label{fig:5a}Particle assembly in the CFD grid}
    \end{subfigure}
    \hfill
    \begin{subfigure}[b]{0.49\textwidth}
        \centering
        \includegraphics[width=5cm,trim=22cm 9cm 30cm 15cm,clip]{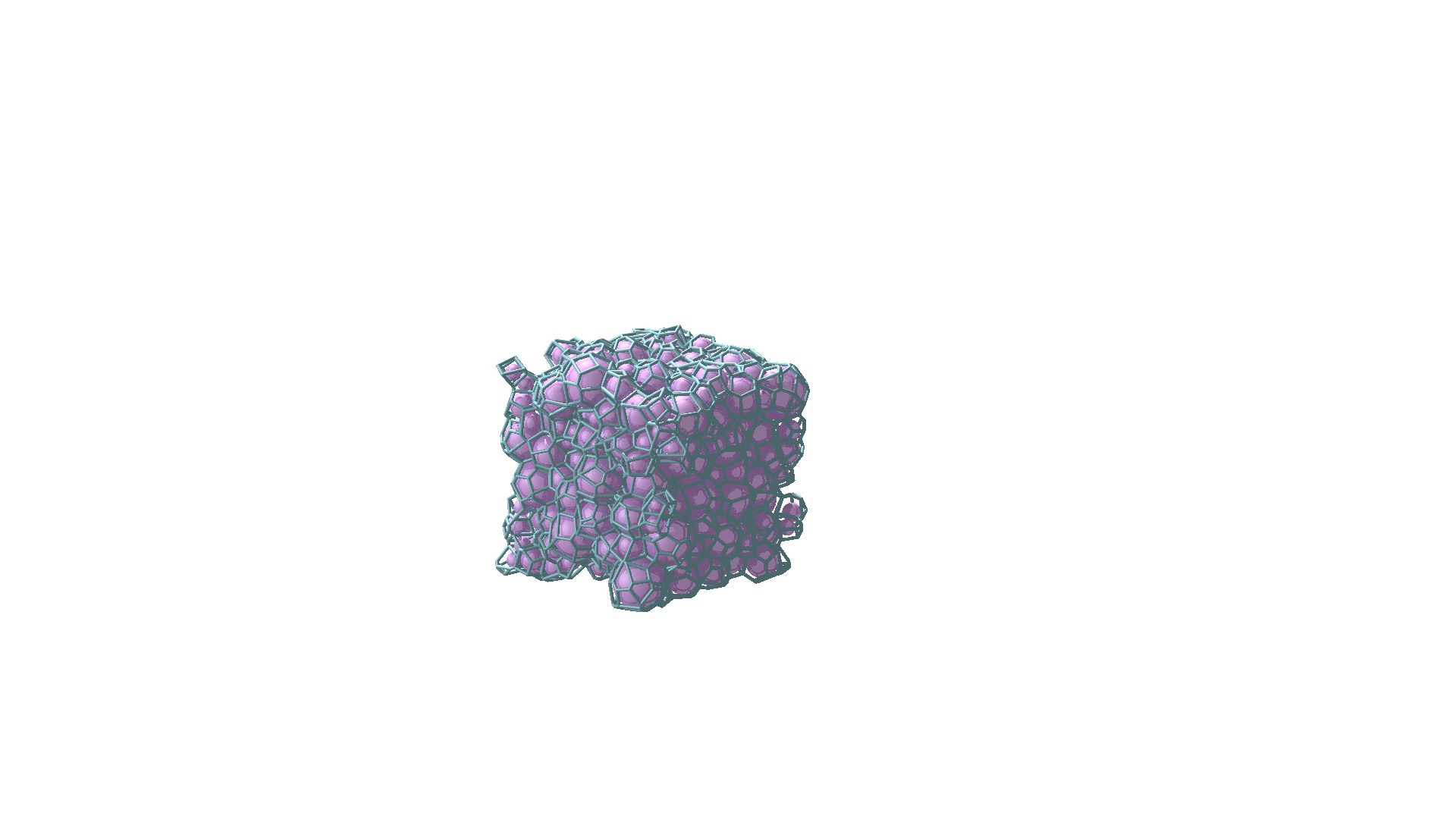}
        \caption{\label{fig:5b}Voronoi tessellation of the particle assembly}
    \end{subfigure}
    \caption{\label{fig:5}Particle assembly in the CFD grid and Voronoi cells.}
\end{figure}

An immersed boundary method (IBM) investigation of fluid-particle interactions in fluid-saturated polydisperse granular materials was carried out by \citet{knight2020}. 
One of the cases considered included 497 particles with a linear grading and the particle diameters range from 0.5 to 1.7 mm. 
To create the samples, particles were randomly placed within cubic periodic boundaries. 
After that, the sample was subject to increasing isotopic compression up to an effective stress of 100 kPa in the DEM solver using servo-controlled periodic boundaries \citep{thornton2000}. 
Six snapshots with global porosity values ranging from 0.319 to 0.602 were taken of the system during compression, and the fluid-particle interactions during laminar flow were determined by IBM simulations using the Multiflow code \citep{denner2014,azis2019}. 
The IBM-generated data is used to validate the proposed 2GVM approach in unresolved CFD-DEM method using the Ergun \citep{ergun1952} and Tenneti \citep{tenneti2011} drag correlations. 
The Ergun and Tenneti drag expressions were chosen as they previously have been applied to consider this system \citep{knight2018}.

\begin{table}[t]
    \caption{\label{tab:1}Particle properties and numerical settings in the simulation Case 1}
    \centering
    \begin{tabular}{c|c|c}
        \hline
        Property & Value & Unit \\
        \hline
        \hline
        Number of particles ($N_p$) & 497 & - \\
        Density ($\rho_p$) & 2470 & kg/m3 \\
        Diameter ($D$) & $0.5 \sim 1.7$ & mm \\
        Coefficient of restitution ($e$) & 0.95 & - \\
        Coulomb friction coefficient ($\mu$) & 0.3 & - \\
        Inlet velocity ($U$) & $2\times10^{-4}$ & m/s \\
        Time step of CFD solver ($\Delta t_1$) & $5\times10^{-6}$ & s \\
        Time step of DEM solver ($\Delta t_2$) & $5\times10^{-8}$ & s \\
        Coupling interval ($M$) & $100 \Delta t_1$ & s \\
        Point cloud density ($\theta_2$) & $0.5 \sim 7.7$ & - \\
        \hline
    \end{tabular}
\end{table}

Fig.~\ref{fig:5} shows the particle assembly along with the CFD grid and Voronoi cells in the coupling. 
In the CFD solver, the boundaries in the $y$ and $z$ direction are periodic. 
In the $x$-direction, the inlet boundary was set to have a fixed fluid velocity and zero pressure gradient, and the outlet was set to a fixed pressure of zero. 
Periodic boundaries are applied in the DEM solver, and the domain for Voronoi tessellation is restricted to the box that is occupied by the particles. 
The sample was fixed in space and not moving with the simulation. 
Table~\ref{tab:1} lists the particle properties and numerical settings in the unresolved CFD-DEM simulations.

\begin{figure}[ht!]
    \centering
    \includegraphics[width=13cm]{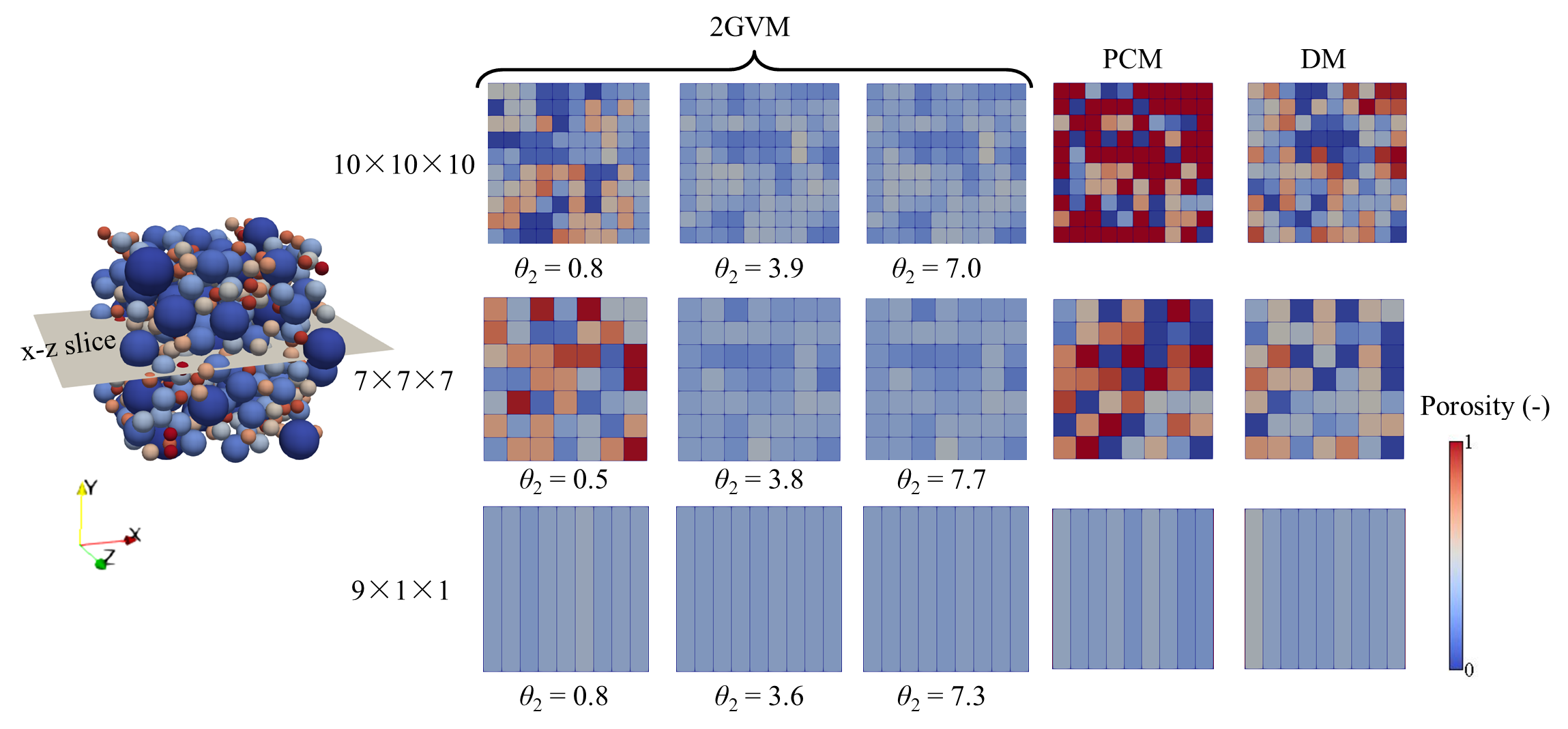}
    \caption{\label{fig:6}Cross-sectional view of the porosity of the particle assembly in the CFD grid.}
\end{figure}

Fig.~\ref{fig:6} shows the porosity distributions on cross-sectional slices in the $x$-$z$ plane for different CFD grid configurations. 
The CFD grid configurations considered are $10\times10\times10$, $7\times7\times7$ and $9\times1\times1$, (number of fluid cells along $x$ $\times$ number of fluid cells along $y$ $\times$ number of fluid cells along $z$). 
These scenarios represent cases where the grid size is slightly smaller than, approximately equal to and larger than the mean particle size of 1.1 mm. 
Results from 2GVM are compared with results obtained using PCM and DM. 
For 2GVM, three different values of $\theta_2$ were selected for each grid configuration to investigate the influence of point cloud density. 

\clearpage

\begin{figure}[ht]
    \centering
    \begin{subfigure}[b]{0.32\textwidth}
        \centering
        \includegraphics[width=5cm]{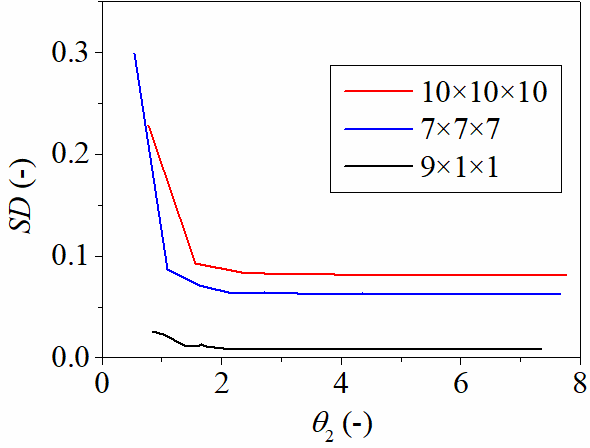}
        \caption{\label{fig:7a}}
    \end{subfigure}
    \hfill
    \begin{subfigure}[b]{0.32\textwidth}
        \centering
        \includegraphics[width=5cm]{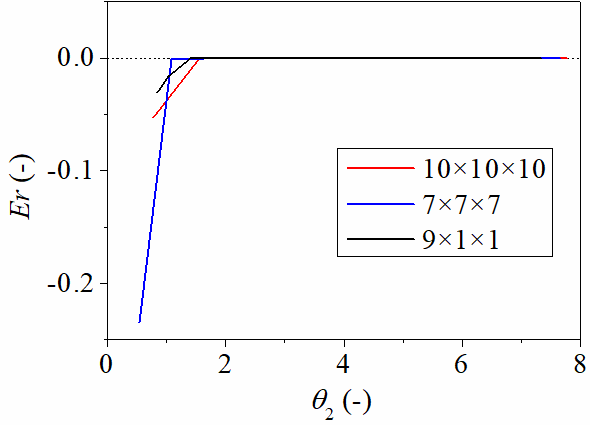}
        \caption{\label{fig:7b}}
    \end{subfigure}
    \hfill
    \begin{subfigure}[b]{0.32\textwidth}
        \centering
        \includegraphics[width=5cm]{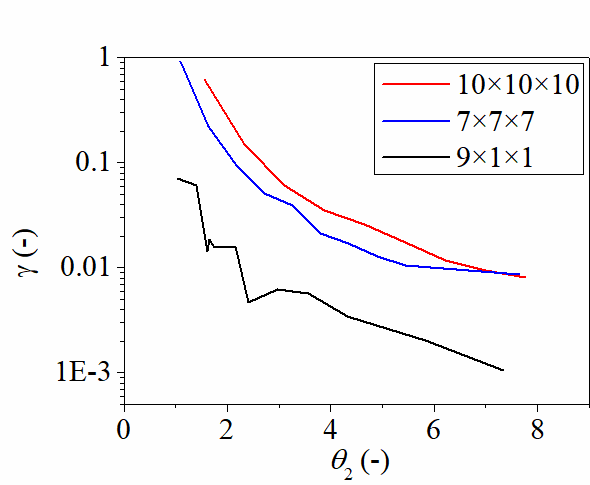}
        \caption{\label{fig:7c}}
    \end{subfigure}
    \caption{\label{fig:7}Variation of $SD$, $Er$ and $\gamma$ with $\theta_2$.}
\end{figure}

While some conclusions can be drawn from Fig.~\ref{fig:6} alone, a comprehensive discussion requires additional analyses. 
Fig.~\ref{fig:7} plots the variation in the $SD$, $Er$ and $\gamma$ with $\theta_2$ for the three CFD grid configurations considered. 
When $\theta_2$ exceeds 3.5, the $SD$ curves converge to specific values, $Er$ converges to zero, and $\gamma$ converge to a value lower than 3\%. 
These data indicate that, for this system, the porosity fields do not measurably change with any further increase in $\theta_2$ for $\theta_2>3.5$. 
All subsequent analysis of this particle assembly is conducted with $\theta_2>3.5$.

\begin{figure}[ht]
    \centering
    \textbf{$\mathbf{10\times10\times10}$ grid:}
    \vfill
    \begin{subfigure}[b]{0.32\textwidth}
        \centering
        \includegraphics[width=5cm]{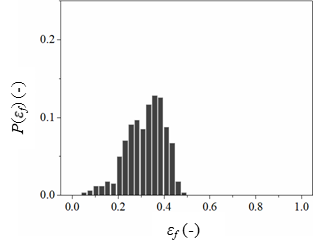}
        \caption{\label{fig:8a}2GVM ($\theta_2 = 3.9$)}
    \end{subfigure}
    \hfill
    \begin{subfigure}[b]{0.32\textwidth}
        \centering
        \includegraphics[width=5cm]{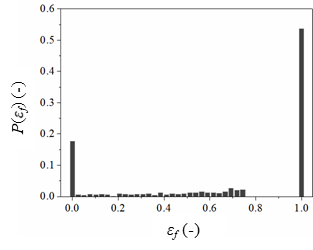}
        \caption{\label{fig:8b}PCM}
    \end{subfigure}
    \hfill
    \begin{subfigure}[b]{0.32\textwidth}
        \centering
        \includegraphics[width=5cm]{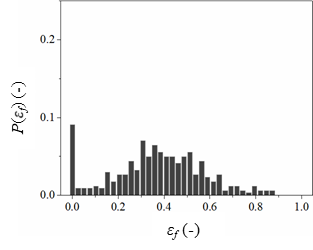}
        \caption{\label{fig:8c}DM}
    \end{subfigure}
    \vfill

    \textbf{$\mathbf{7\times7\times7}$ grid:}
    \vfill
    \centering
    \begin{subfigure}[b]{0.32\textwidth}
        \centering
        \includegraphics[width=5cm]{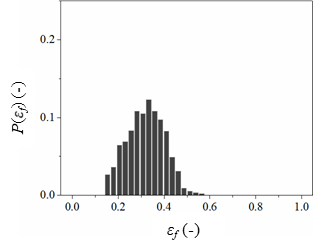}
        \caption{\label{fig:8d}2GVM ($\theta_2 = 3.8$)}
    \end{subfigure}
    \hfill
    \begin{subfigure}[b]{0.32\textwidth}
        \centering
        \includegraphics[width=5cm]{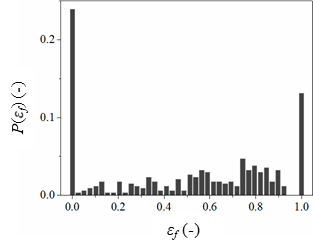}
        \caption{\label{fig:8e}PCM}
    \end{subfigure}
    \hfill
    \begin{subfigure}[b]{0.32\textwidth}
        \centering
        \includegraphics[width=5cm]{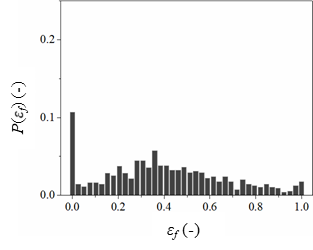}
        \caption{\label{fig:8f}DM}
    \end{subfigure}
    \caption{\label{fig:8}Histograms of the porosity from different coarse-graining methods.}
\end{figure}

Fig.~\ref{fig:8} shows the histogram of porosity values obtained from the 2GVM (with $\theta_2>3.5$), PCM and DM. 
Consideration of the combined data on Fig.~\ref{fig:6} and Fig.~\ref{fig:8} shows that the porosity values obtained from the PCM and DM have non-physical porosity values (approximating 0 or 1.0) and large jumps between the neighbouring cells in the $10\times10\times10$ and $7\times7\times7$ grid cases. 
The non-physical features are not observed in the 2GVM approach and the porosity field becomes smooth, which is consistent with the uniformity of the particle assembly in this case. 
For the one-dimensional (1D) case ($9\times1\times1$ grid), each grid cell contains around 42 particles, the effects of coarse-graining approaches on the porosity field are not obvious.

\begin{figure}[ht]
    \centering
    \includegraphics[width=10cm]{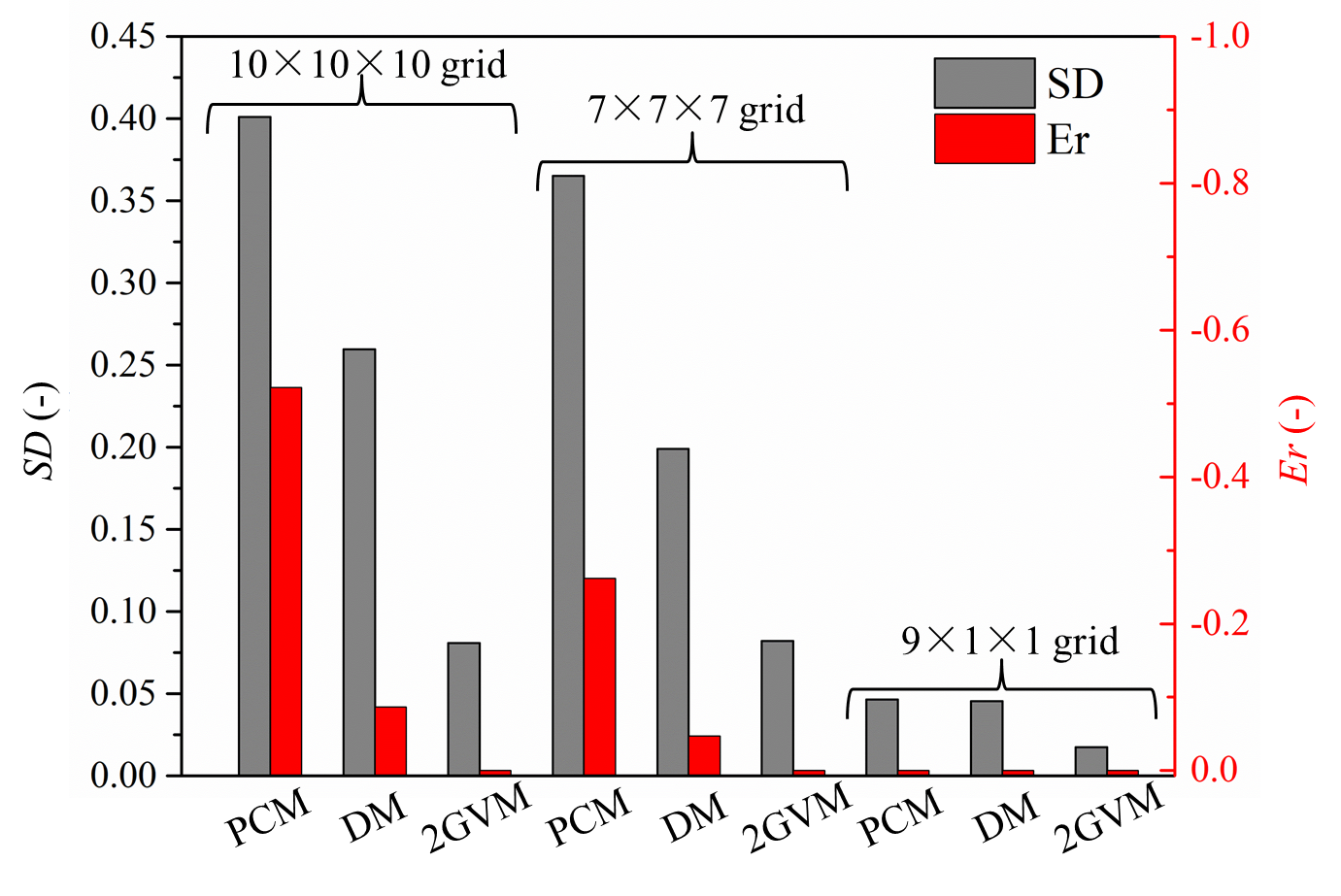}
    \caption{\label{fig:9}Comparison of the $SD$ of the porosity values and the $Er$ of solid volume for difference CFD grid configurations and coarse-graining methods.}
\end{figure}

\begin{figure}[ht]
    \centering
    \begin{subfigure}[b]{0.32\textwidth}
        \centering
        \includegraphics[width=5cm]{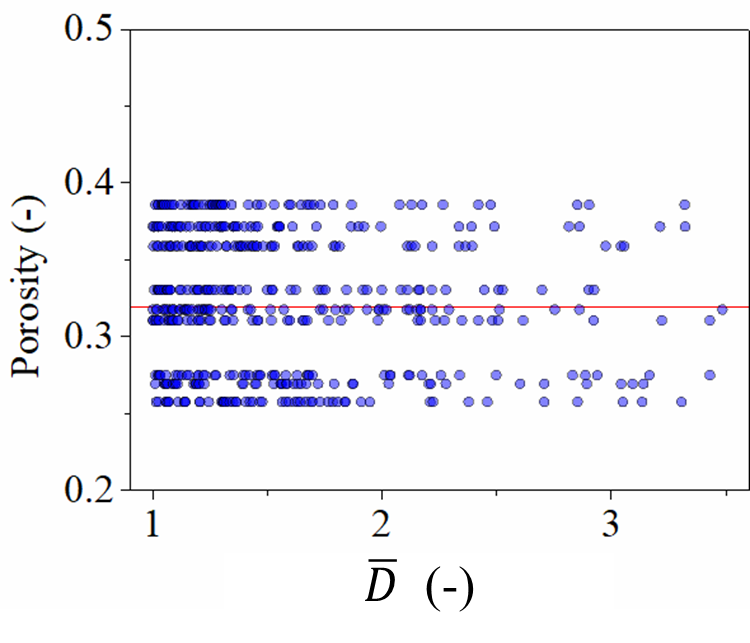}
        \caption{\label{fig:10a}PCM}
    \end{subfigure}
    \hfill
    \begin{subfigure}[b]{0.32\textwidth}
        \centering
        \includegraphics[width=5cm]{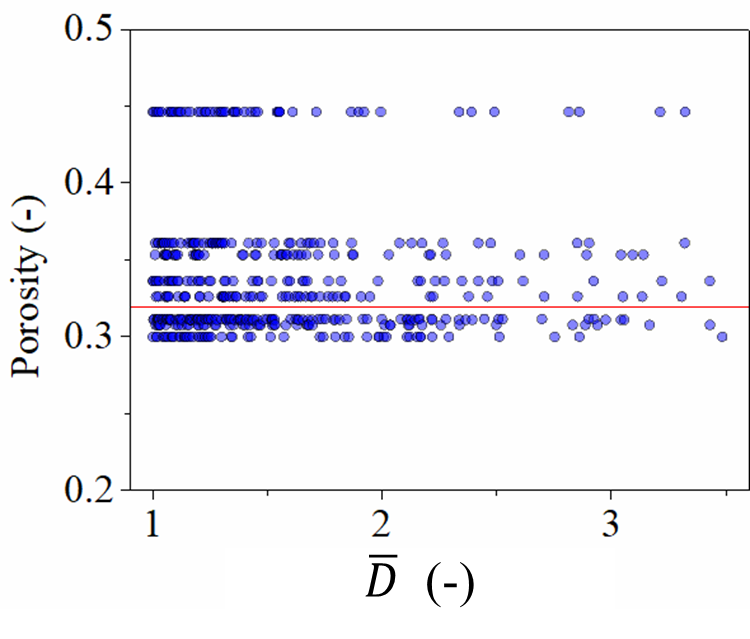}
        \caption{\label{fig:10b}DM}
    \end{subfigure}
    \hfill
    \begin{subfigure}[b]{0.32\textwidth}
        \centering
        \includegraphics[width=5cm]{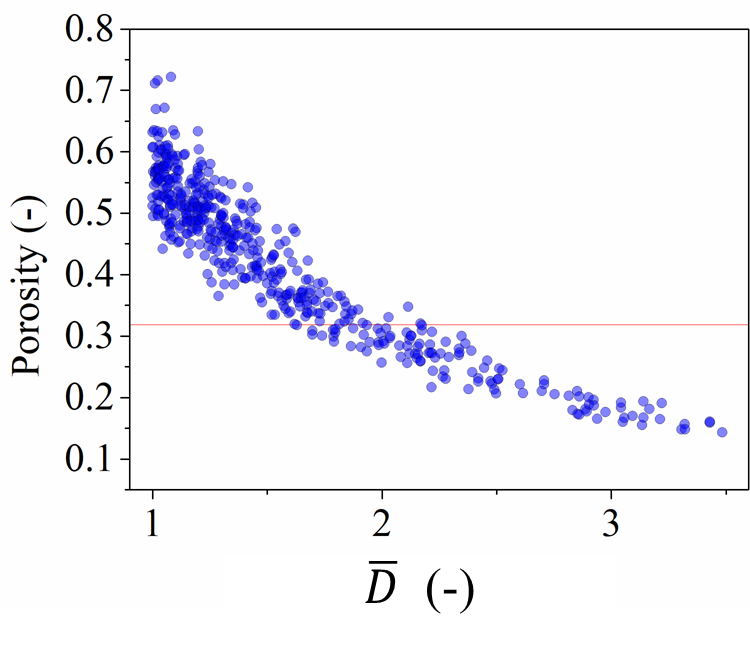}
        \caption{\label{fig:10c}2GVM}
    \end{subfigure}
    \caption{\label{fig:10}Local particle porosity against the normalized particle diameter for different coarse-graining methods. The global porosity is indicated by the red line.}
\end{figure}

Fig.~\ref{fig:9} provides bar charts illustrating the $SD$ of the porosity distributions and the $Er$ of $V^C_{p,total}$ to the real solid volume. 
When coarse-graining, the porosity of a CFD cell must be between 0 and 1. 
However, in both PCM and DM, the particle volume can exceed the the CFD cell volume if the particle is larger than the CFD cell. 
In these instances, the excess volume must be discarded to ensure that the porosity lies between 0 and 1, resulting in an error in the total particle volume exchanged with the CFD solver. 
Fig.~\ref{fig:9} shows that 2GVM gives the lowest SD and Er values amongst the three approaches, and thus the porosity field generated by the 2GVM is the most uniform and best able to conserve volume. 
On the other hand, PCM and DM provide results in which volume is conserved only in the 1D case, in which each CFD grid cell contains more than 40 particles.

In Fig.~\ref{fig:10}, the local particle porosities calculated from the three coarse-graining methods are plotted against the normalized diameters for the case where the global porosity is 0.319. 
It is clear that as the particle diameter increases the porosity calculated by the 2GVM decreases significantly, which is to be expected as (relatively) larger particles are known to experience greater local solid fractions \citep{knight2020}. 
On the other hand, all the porosities from the PCM and DM are assigned to one of a finite number of fixed levels. 
This is because each particle is assigned the same porosity value of the specific CFD grid cell in which it is positioned and each cell contains multiple particles.

\clearpage

\begin{table}[ht]
    \caption{\label{tab:2}Pearson correlation between the individual particle drag forces from the IBM and unresolved CFD-DEM cases. The global $\varepsilon_f = 0.319$.}
    \centering
    \begin{tabular}{c|c|c|c}
        \hline
        Drag correlation & PCM & DM & 2GVM \\
        \hline
        \hline
        Ergun & 0.61 & 0.78 & 0.82 \\
        Tenneti & 0.62 & 0.78 & 0.81 \\
        \hline
    \end{tabular}
\end{table}

\begin{figure}[ht!]
    \centering
    \begin{subfigure}[b]{0.49\textwidth}
        \centering
        \includegraphics[width=5cm]{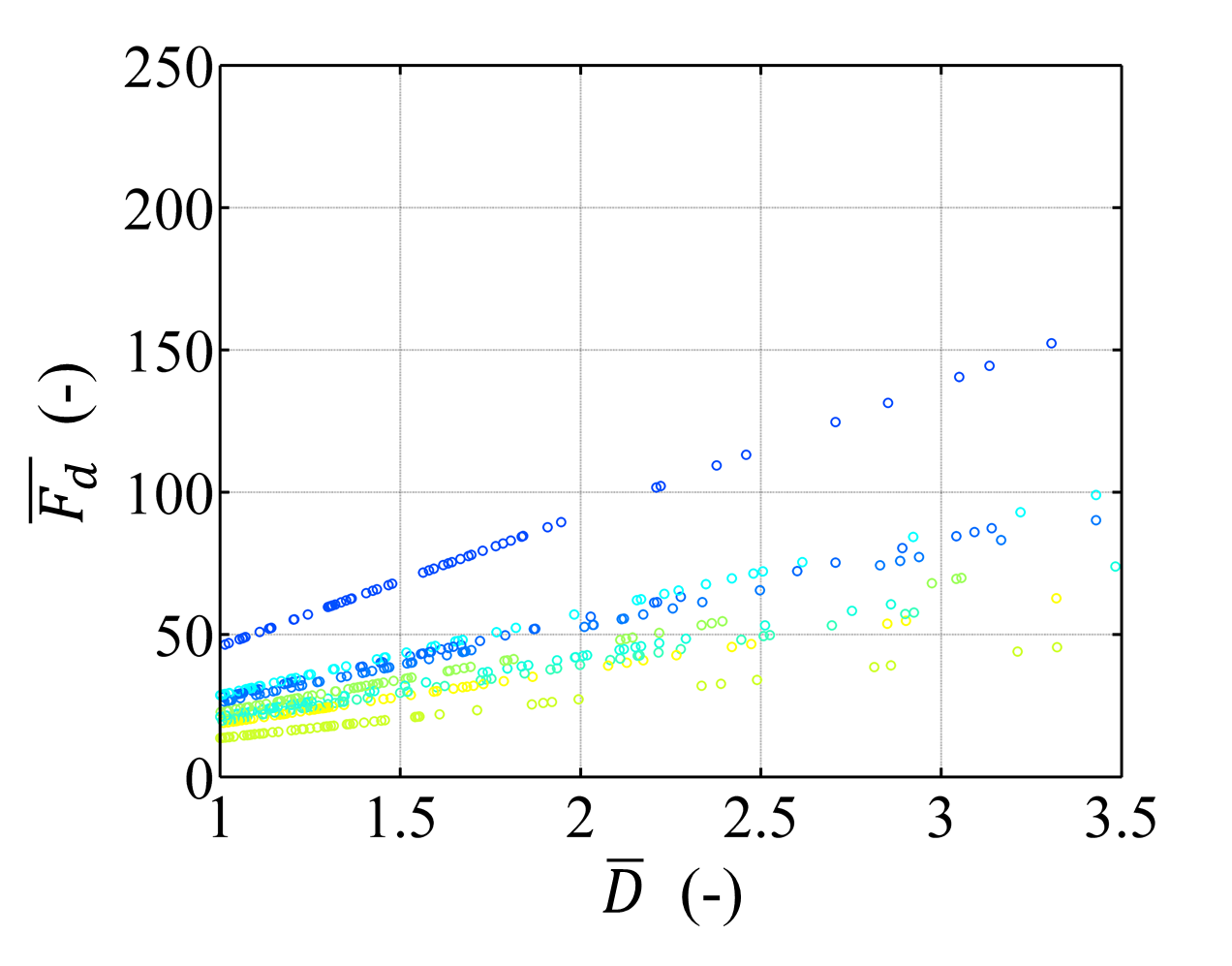}
        \caption{\label{fig:11a}PCM, Ergun}
    \end{subfigure}
    \hfill
    \begin{subfigure}[b]{0.49\textwidth}
        \centering
        \includegraphics[width=5cm]{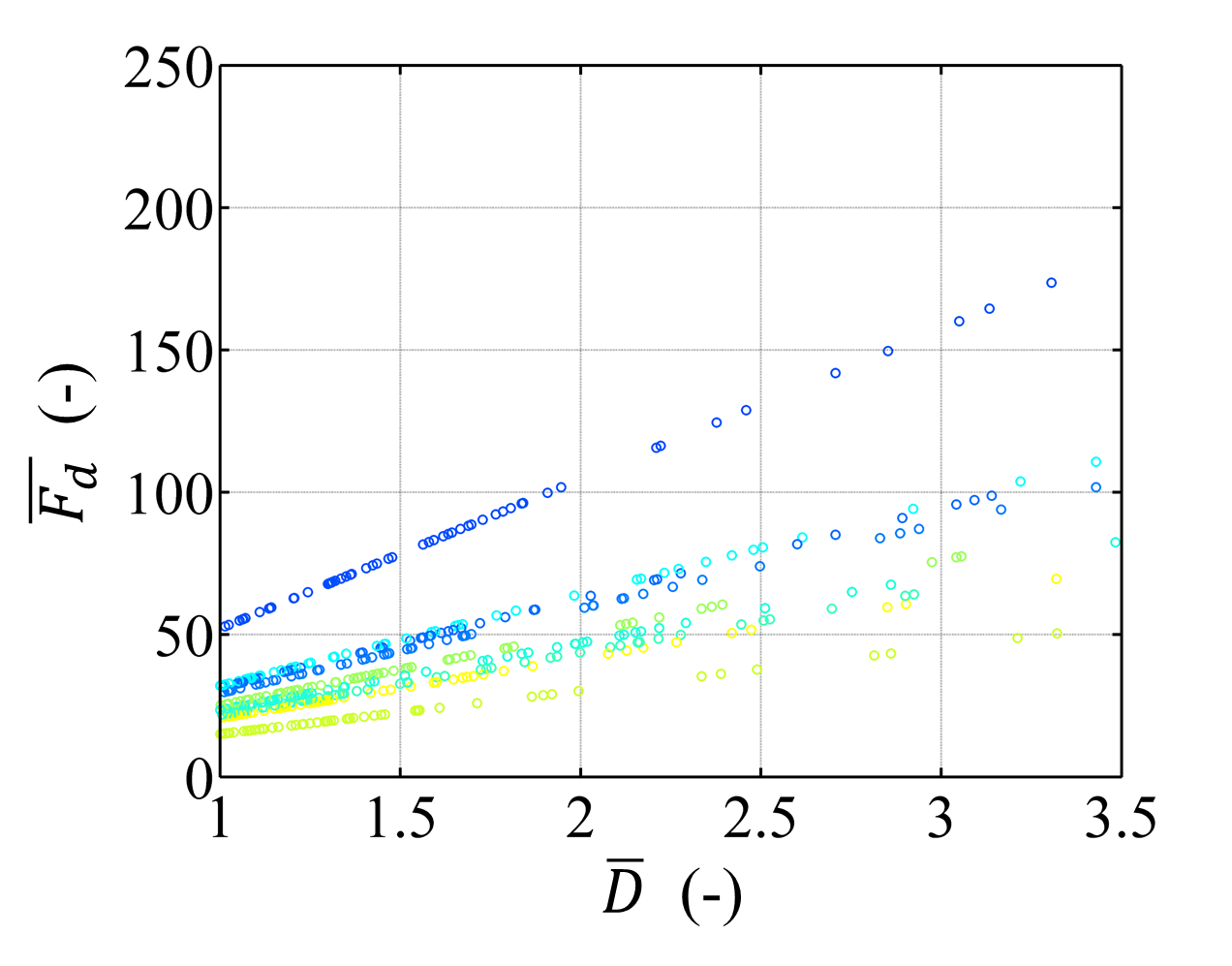}
        \includegraphics[height=4cm]{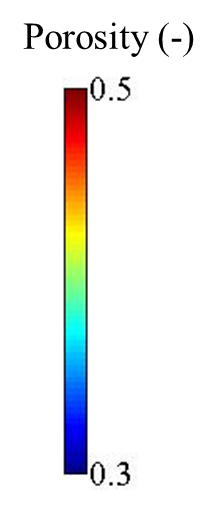}
        \caption{\label{fig:11b}PCM, Tenneti}
    \end{subfigure}
    \vfill
    \begin{subfigure}[b]{0.49\textwidth}
        \centering
        \includegraphics[width=5cm]{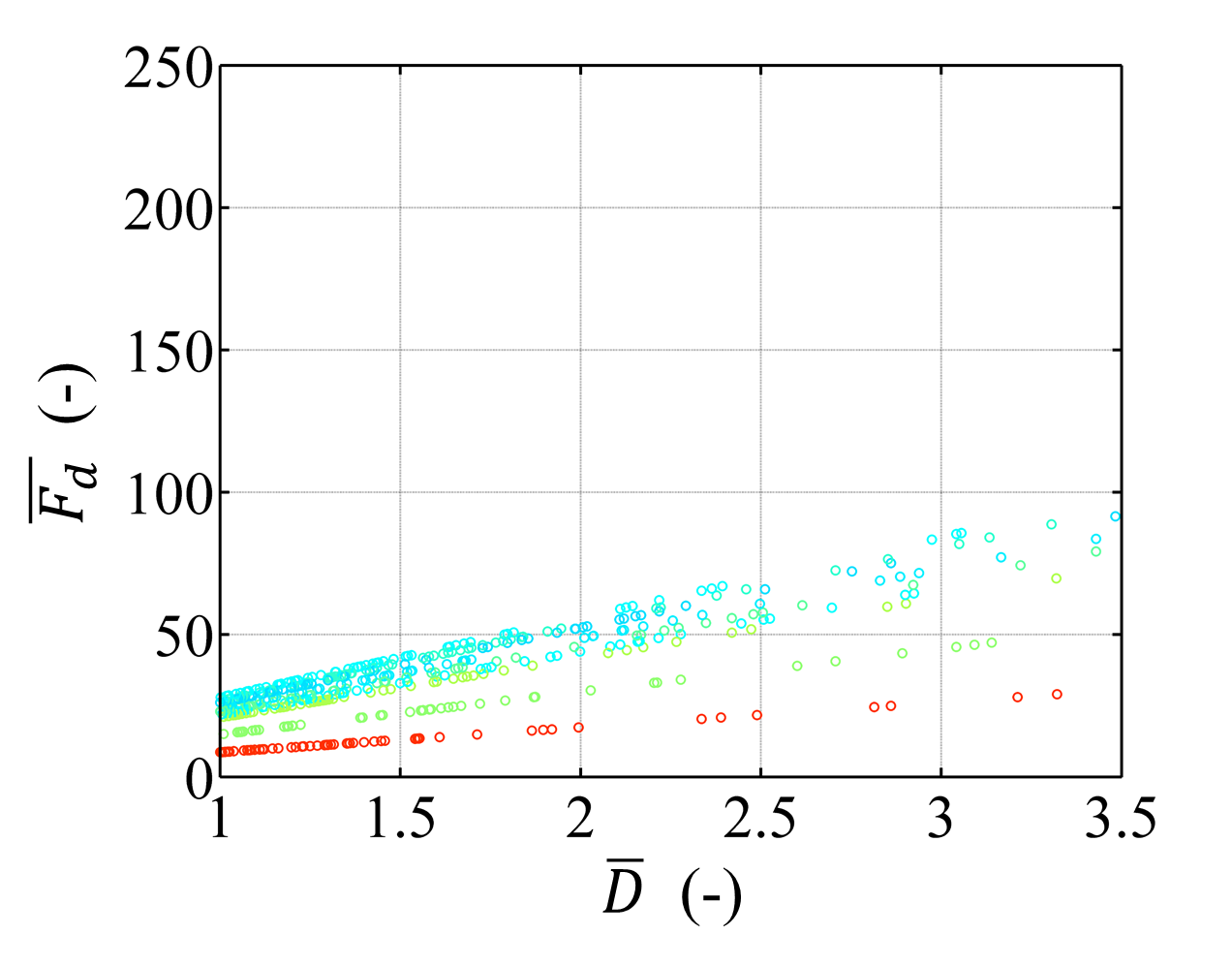}
        \caption{\label{fig:11d}DM, Ergun}
    \end{subfigure}
    \hfill
    \begin{subfigure}[b]{0.49\textwidth}
        \centering
        \includegraphics[width=5cm]{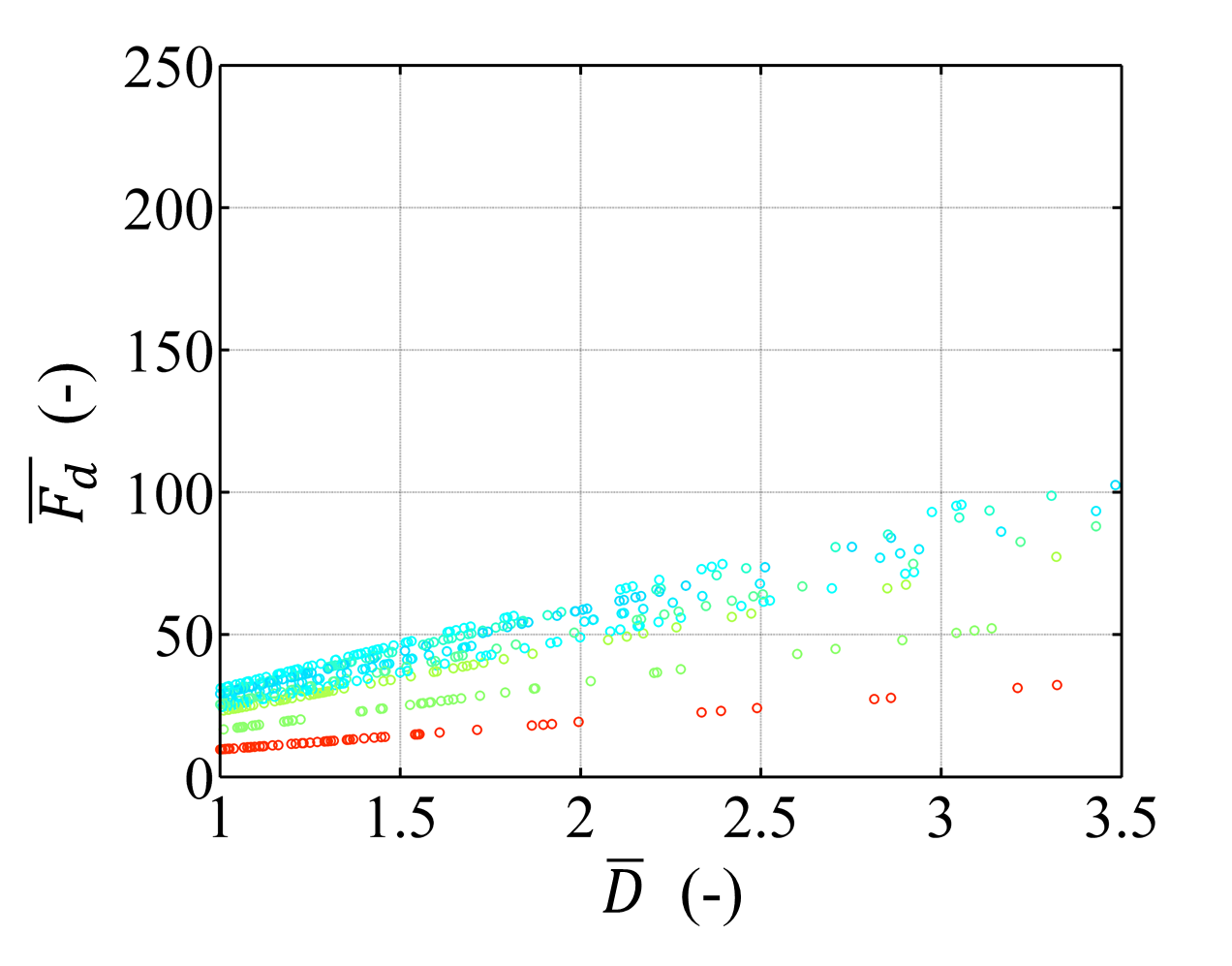}
        \includegraphics[height=4cm]{11c.png}
        \caption{\label{fig:11e}DM, Tenneti}
    \end{subfigure}
    \vfill
    \begin{subfigure}[b]{0.49\textwidth}
        \centering
        \includegraphics[width=5cm]{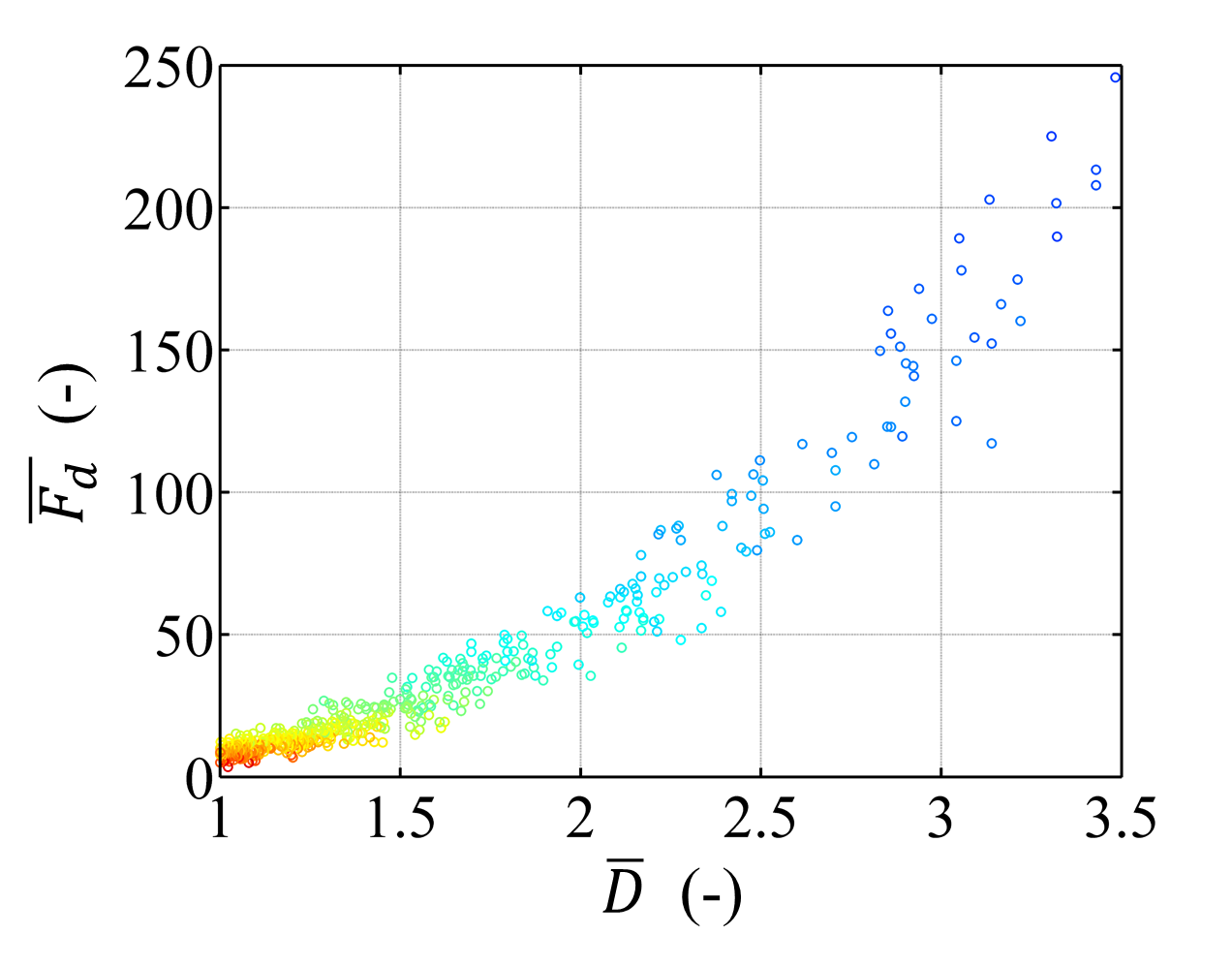}
        \caption{\label{fig:11f}2GVM, Ergun}
    \end{subfigure}
    \hfill
    \begin{subfigure}[b]{0.49\textwidth}
        \centering
        \includegraphics[width=5cm]{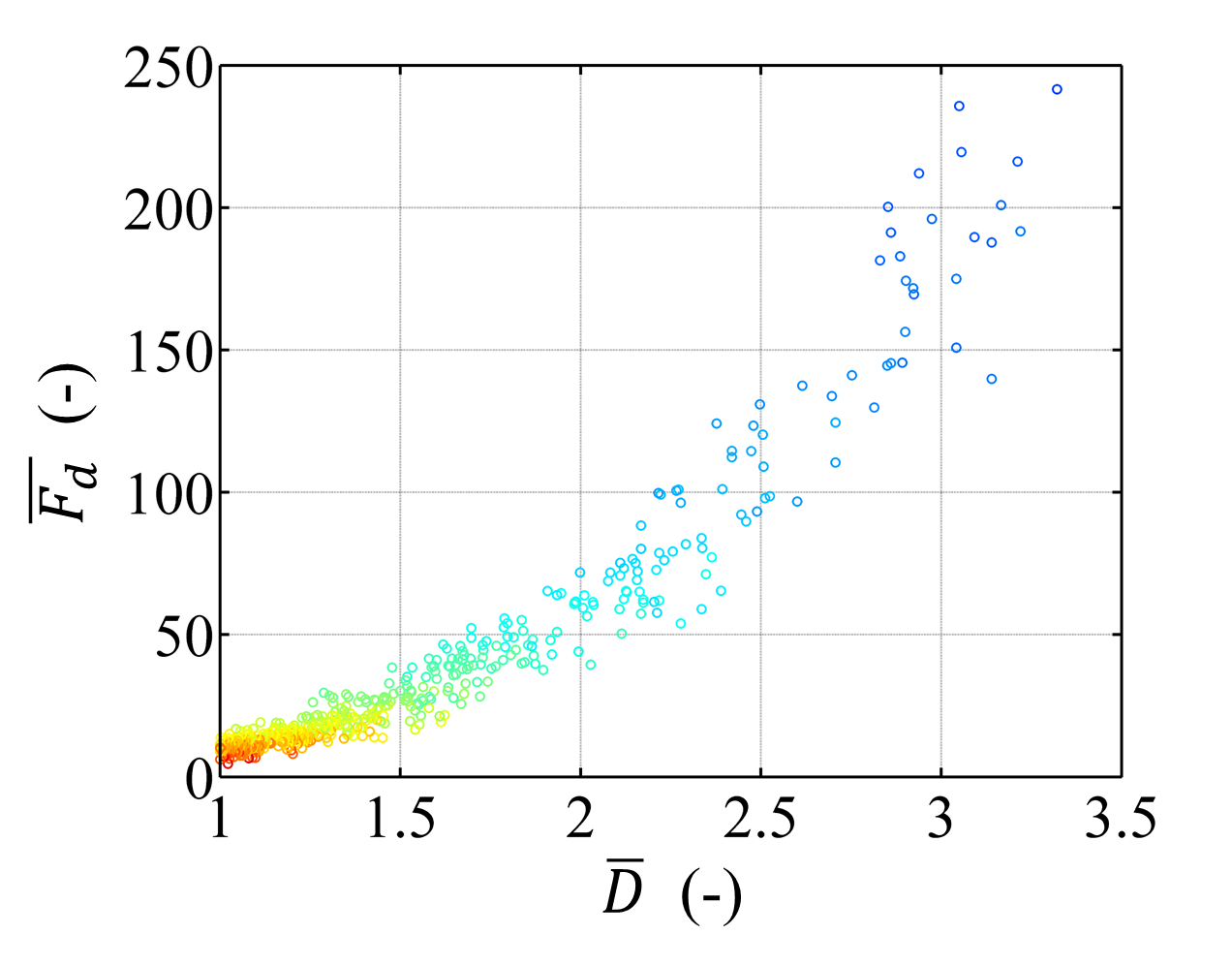}
        \includegraphics[height=4cm]{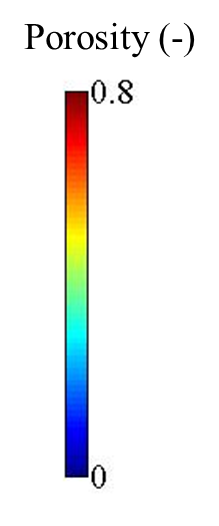}
        \caption{\label{fig:11g}2GVM, Tenneti}
    \end{subfigure}
    \caption{\label{fig:11}Individual particle drag forces against diameter at global $\varepsilon_f = 0.319$ with the drag forces extracted according to PCM, DM and 2GVM ($9\times1\times1$ CFD grid, $\theta_2 = 3.6$).}
\end{figure}

\begin{figure}[ht]
    \centering
    \begin{subfigure}[b]{0.32\textwidth}
        \centering
        \includegraphics[width=5cm]{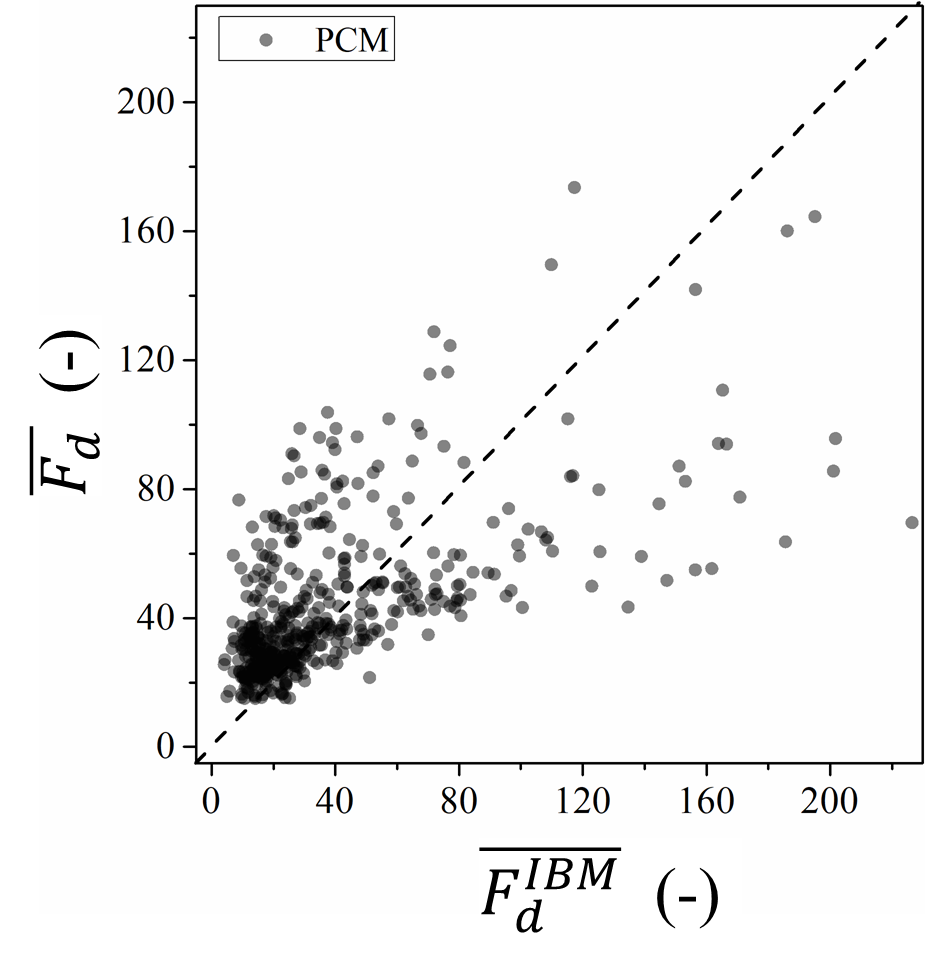}
        \caption{\label{fig:12a}PCM, Ergun}
    \end{subfigure}
    \hfill
    \begin{subfigure}[b]{0.32\textwidth}
        \centering
        \includegraphics[width=5cm]{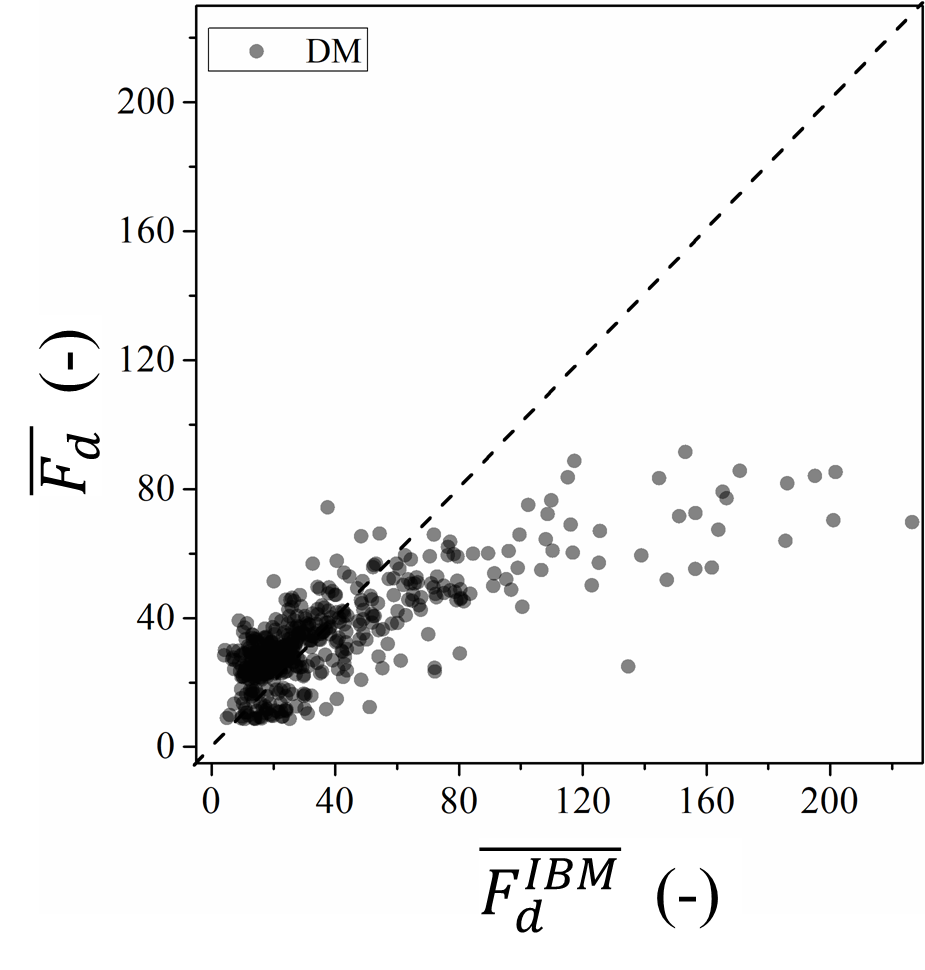}
        \caption{\label{fig:12b}DM, Ergun}
    \end{subfigure}
    \hfill
    \begin{subfigure}[b]{0.32\textwidth}
        \centering
        \includegraphics[width=5cm]{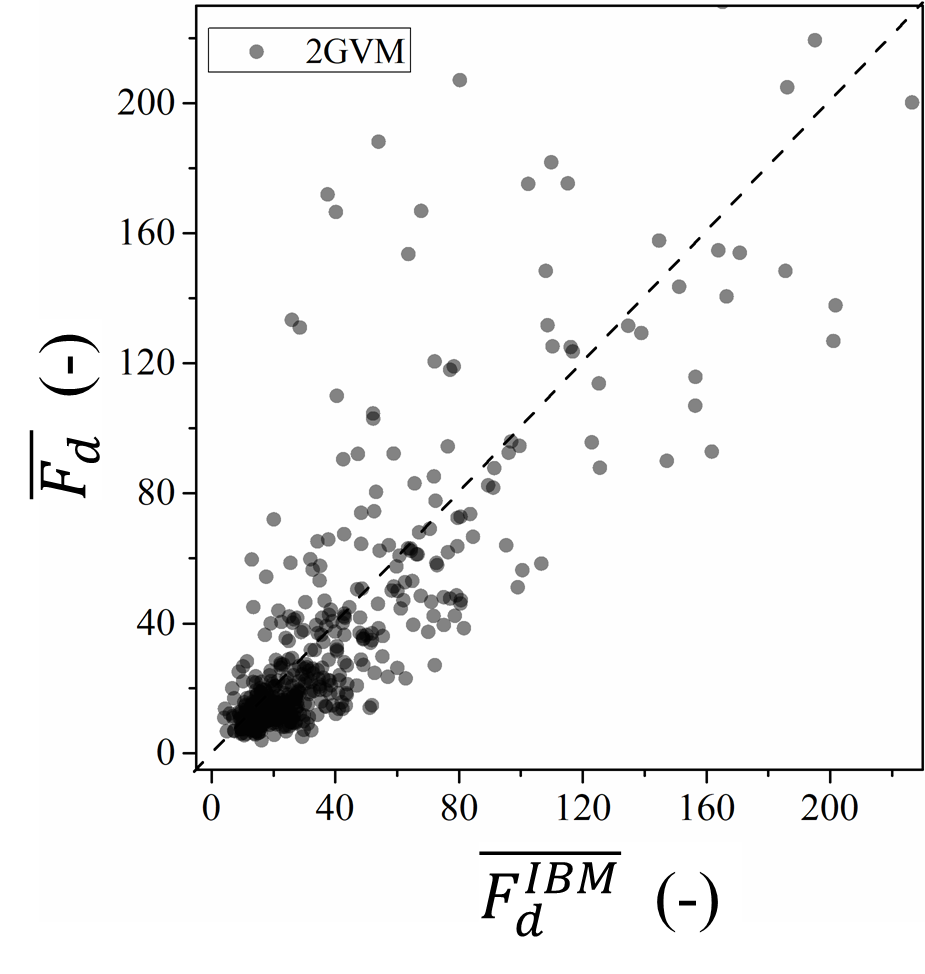}
        \caption{\label{fig:12c}2GVM, Ergun}
    \end{subfigure}
    \vfill
    \begin{subfigure}[b]{0.32\textwidth}
        \centering
        \includegraphics[width=5cm]{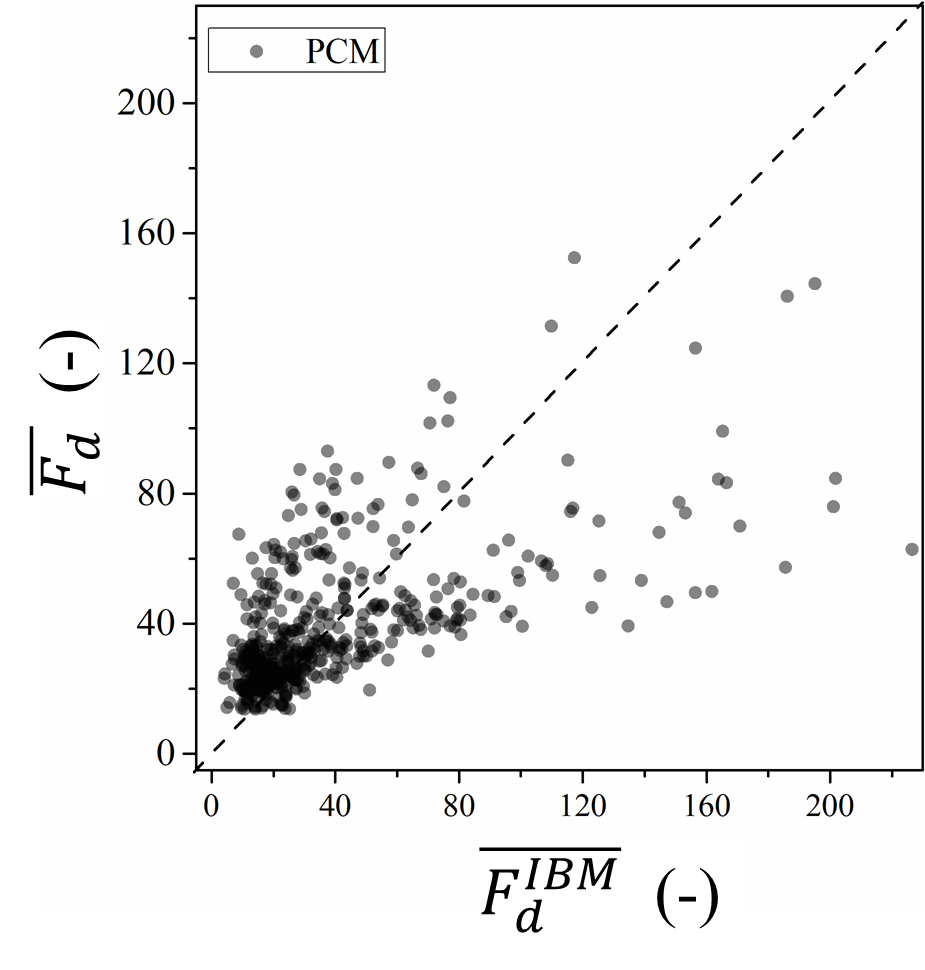}
        \caption{\label{fig:12d}PCM, Tenneti}
    \end{subfigure}
    \hfill
    \begin{subfigure}[b]{0.32\textwidth}
        \centering
        \includegraphics[width=5cm]{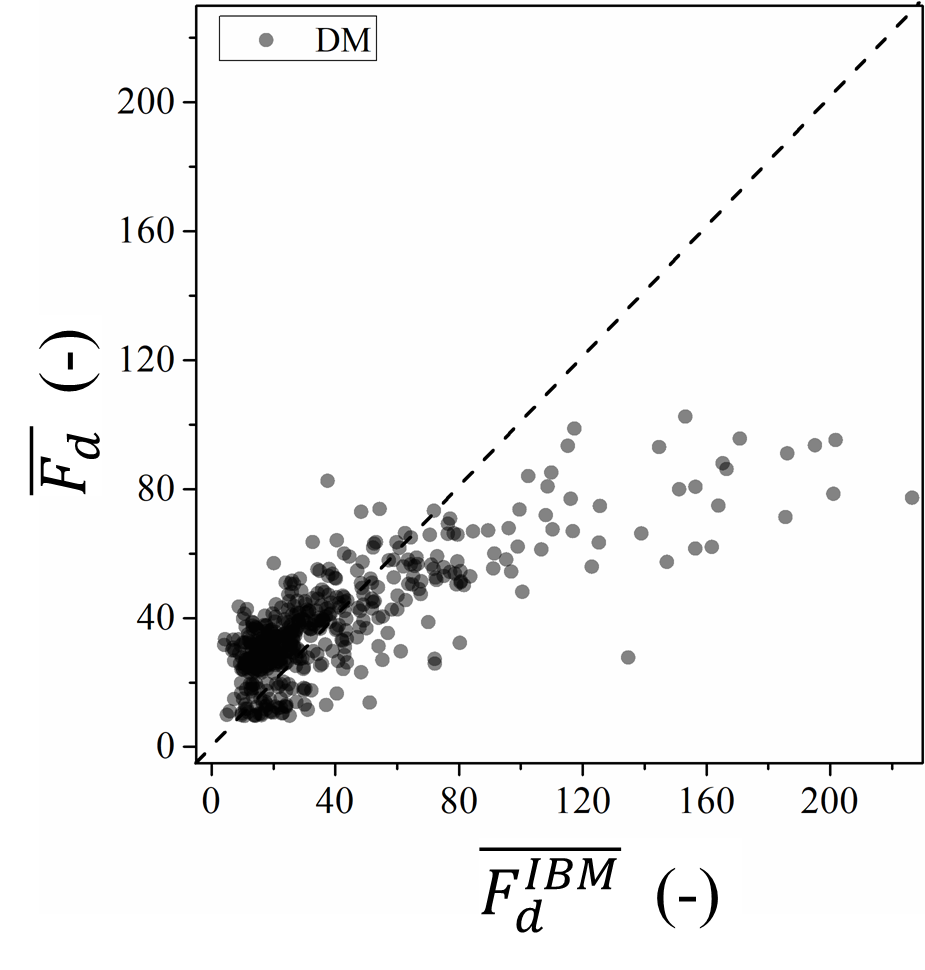}
        \caption{\label{fig:12e}DM, Tenneti}
    \end{subfigure}
    \hfill
    \begin{subfigure}[b]{0.32\textwidth}
        \centering
        \includegraphics[width=5cm]{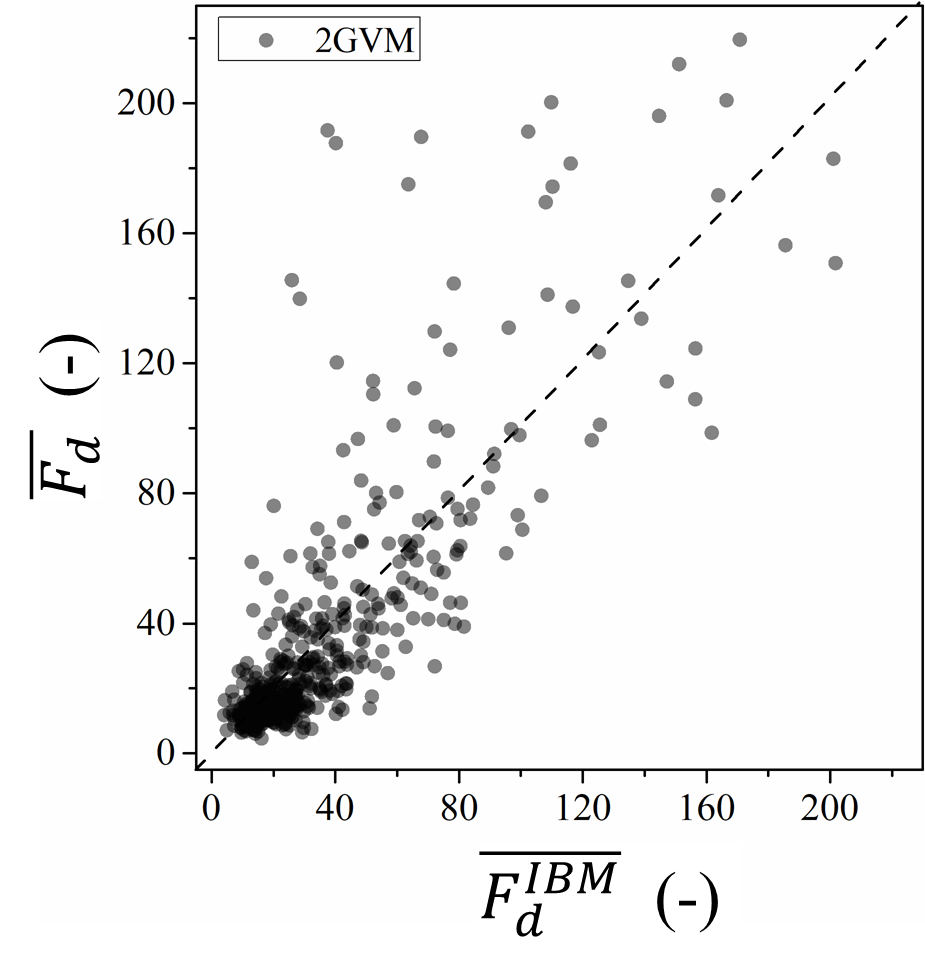}
        \caption{\label{fig:12f}2GVM, Tenneti}
    \end{subfigure}
    \hfill
    \caption{\label{fig:12}Individual particle drag forces against those from IBM simulations. The global $\varepsilon_f = 0.319$. Note that the data points are transparent – the darker shading indicates overlapping data points.}
\end{figure}

Due to the small number of particles in the sample, simulation cases with the PCM and DM could only be reasonably run using the 1D CFD grid ($9\times1\times1$). 
For this case, the magnitude of the drag force acting on each particle normalized by the Stokes force ($\overline{F_d}$) is plotted against normalised particle diameter ($\overline{D}$) in Fig.~\ref{fig:11}. 
The local porosity is indicated by the colour of the data points. 
In Fig.~\ref{fig:11}, the individual particle drag forces calculated using the PCM and DM approaches are distributed along several straight lines. 
Each line represents data for particles with the same porosity values (in the same CFD cell). 
The reason the correlations between $\overline{F_d}$ and $\overline{D}$ are linear can be understood by reference to the analytical forms of the normalised drag forces provided in \citet{knight2018}. 
At a given porosity, the $\overline{F_d}$ calculated from the Ergun correlation is linearly proportional to the Reynolds number which in turn is linearly dependant on the particle diameter. 
The normalised Tenneti drag expression is more complex involving $Re^{0.687}$ and $Re$ terms, and so $\overline{F_d}$ is approximately linearly dependent on the particle diameter when 0.10 < $Re$ < 0.33. 
In contrast, the drag forces calculated from 2GVM show a different, nonlinear trend, and are clearly dependent on the particle diameter. 
The Tenneti correlation provides slightly greater drag forces than those predicted using the Ergun correlation. 
The individual particle drag forces are plotted against the drag forces obtained from IBM simulations \citep{knight2020} considering the same particle assembly in Fig.~\ref{fig:12}, the corresponding Pearson correlation values are list in Table~\ref{tab:2}. 
The data on Fig.~\ref{fig:12} and Table~\ref{tab:2} clearly show that the drag force predictions obtained using the 2GVM more closely follow the trends observed in the IBM data than the predicted values obtained using PCM and DM. 
Since PCM and DM neglect the effect of particle size on the local particle-scale porosity, the predictions from these methods slightly overestimate the drag force acting on small particles and underestimate the drags for large particles. 
This is significant as any relative movement between particles may be underestimated. 
In unresolved CFD-DEM, the fluid flow around an individual particle is not accurately simulated and the drag force is estimated by empirical models. 
Thus, it is not expected to achieve results that exactly agree with the IBM. 
However, these data indicate that a significant improvement in the prediction of the drag force on the individual particle can be realised by applying the 2GVM with a relative computational cost significantly lower than required for resolved CFD-DEM approaches such as the IBM.

\begin{figure}[ht!]
    \centering
    \begin{subfigure}[b]{0.49\textwidth}
        \centering
        \includegraphics[width=6cm]{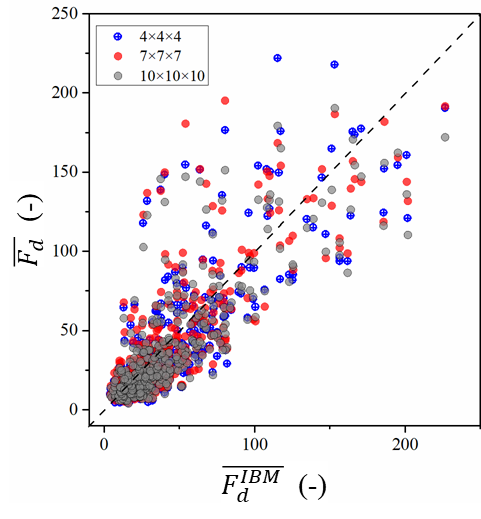}
        \caption{\label{fig:13a}$\varepsilon_f = 0.319$}
    \end{subfigure}
    \hfill
    \begin{subfigure}[b]{0.49\textwidth}
        \centering
        \includegraphics[width=6cm]{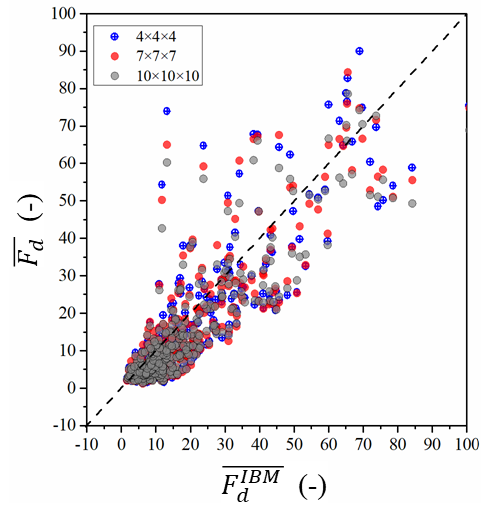}
        \caption{\label{fig:13b}$\varepsilon_f = 0.478$}
    \end{subfigure}
    \vfill
    \begin{subfigure}[b]{0.49\textwidth}
        \centering
        \includegraphics[width=6cm]{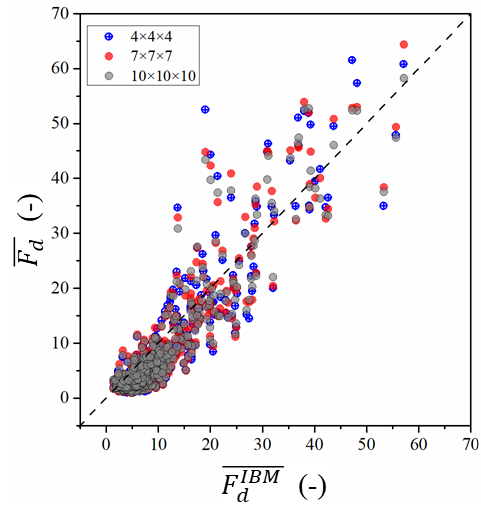}
        \caption{\label{fig:13c}$\varepsilon_f = 0.551$}
    \end{subfigure}
    \hfill
    \begin{subfigure}[b]{0.49\textwidth}
        \centering
        \includegraphics[width=6cm]{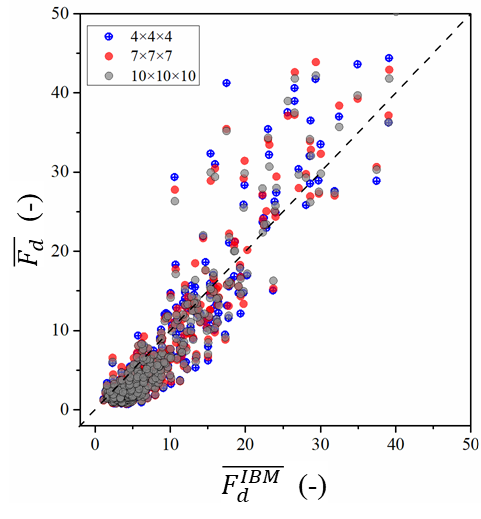}
        \caption{\label{fig:13d}$\varepsilon_f = 0.602$}
    \end{subfigure}
    \caption{\label{fig:13}Local particle drag forces from 2GVM with different mesh densities against the local particle drag forces from IBM simulations. }
\end{figure}

Fig.~\ref{fig:13} considers the sensitivity of the performance of 2GVM to packing density and grid configuration,  Table~\ref{tab:3} lists the corresponding Pearson correlation between the data from 2GVM and IBM. 
The individual particle drag forces calculated using 2GVM are plotted against the local particle drag forces from IBM simulations for 4 porosity values ranging from 0.319 to 0.602. 
For each porosity considered, data obtained with grid densities of $4\times4\times4$, $7\times7\times7$ and $10\times10\times10$ are compared. 
In all cases the Ergun drag correlation was applied. 
Relatively low Pearson correlation values can be seen in the cases with low global porosities, which indicates the prediction is less accurate in the case of densely packed particle assemblies. 
This can most likely be attributed to the fact that the overall porosity of densely packed case is slightly out of the recommended scope of the Ergun ($0.320 \le \varepsilon_f \le 0.470$) and Tenneti ($0.350\le \varepsilon_f \le0.90$) expressions which have an empirical basis \citep{knight2018}. 
For cases with the same global porosity, the distribution of the drag force from different CFD grid densities are in close agreement, thus data from CFD-DEM simulations using the 2GVM are relatively independent of the CFD grid density. 

\begin{table}[ht]
    \caption{\label{tab:3}Pearson correlation between the individual particle drag force from IBM and unresolved CFD-DEM.}
    \centering
    \begin{tabular}{c|c|c|c|c}
        \hline
        Grid Density & $\varepsilon_f = 0.319$ & $\varepsilon_f = 0.478$ & $\varepsilon_f = 0.551$ & $\varepsilon_f = 0.602$ \\
        \hline
        \hline
        $4\times4\times4$       & 0.83 & 0.86 & 0.91 & 0.92 \\
        $7\times7\times7$       & 0.83 & 0.87 & 0.92 & 0.93 \\
        $10\times10\times10$    & 0.84 & 0.88 & 0.92 & 0.93 \\
        \hline
    \end{tabular}
\end{table}

\begin{figure}[ht]
    \centering
    \includegraphics[width=10cm]{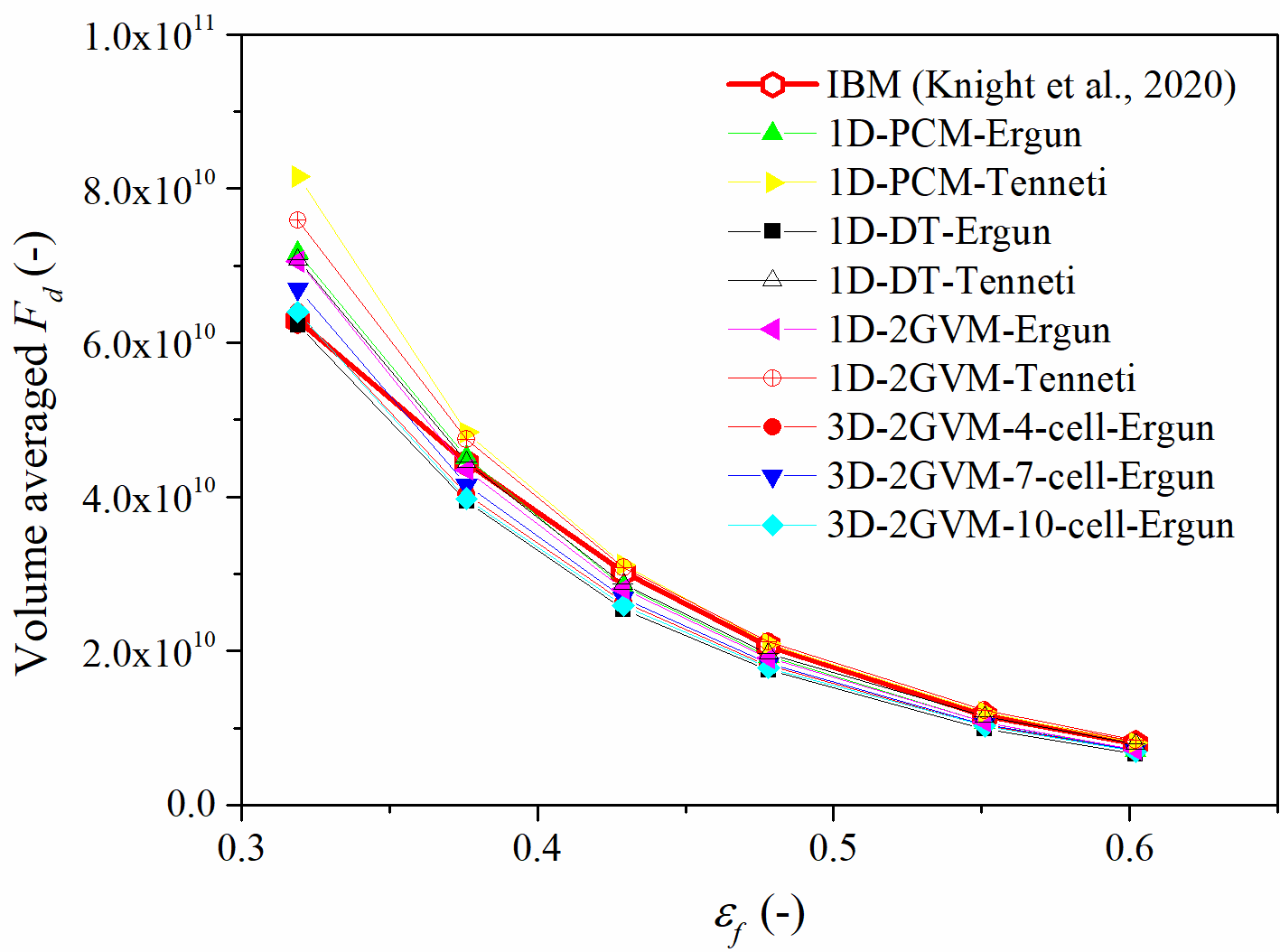}
    \caption{\label{fig:14}Variation in volume averaged drag force with global porosity.}
\end{figure}

The overall drag force applied to the assembly was divided by the total volume of the assembly to obtain the volume-averaged drag force which is shown in Fig.~\ref{fig:14}. 
The results calculated from different approaches are in a close agreement. Combing Fig.~\ref{fig:11}, Fig.~\ref{fig:12} and Fig.~\ref{fig:14}, it can be concluded that the 2GVM provides excellent prediction in the individual particle drag force and the overall drag force of the particle assembly. 
The PCM and DM provide a good prediction of the overall drag force of the particle assembly even though the data presented above showed they are unable to predict individual particle drag forces correctly. 
Therefore, 2GVM is a better choice in the case of polydisperse systems, to ensure that relative particle motion is accurately captured. 

\subsection{\label{sub:4.2}Validation Case 2: Spouted fluidised bed}

\begin{figure}[ht]
    \centering
    \includegraphics[width=6cm]{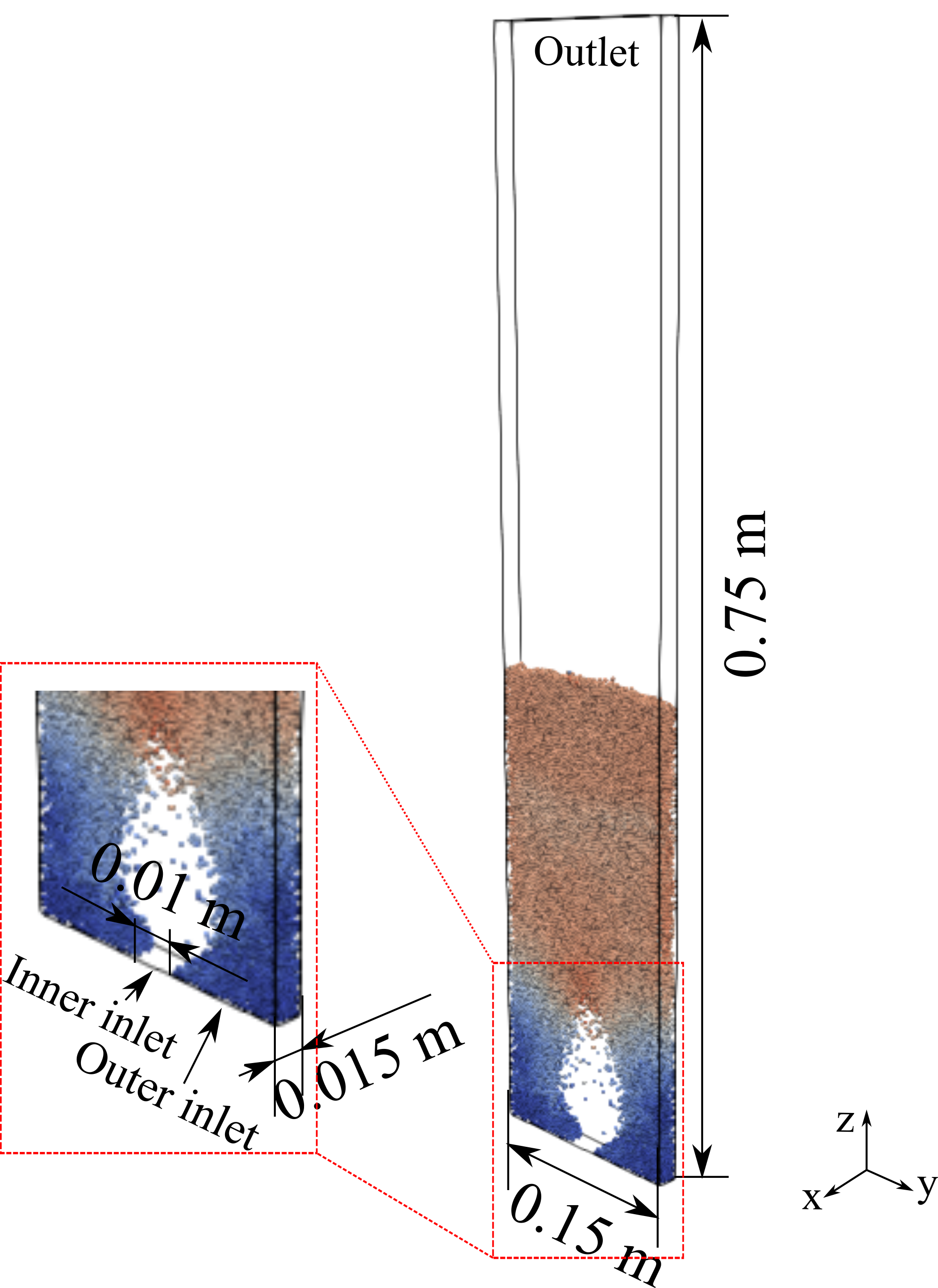}
    \caption{\label{fig:15}The dimensions and grid of the spouted fluidised bed.}
\end{figure}

To evaluate the ability of 2GVM to predict the dynamic mechanisms associated with fluidisation processes, the second validation case considered published experimental data for a pseudo-2D spouted fluidised bed documented in \citet{link2005}. 
Fig.~\ref{fig:15} shows the dimensions of the targeted spouted fluidised bed, which has two inlets and one outlet. 
The inner inlet and outer inlet introduce the upward fluidisation air for spout fluidisation and background fluidisation, respectively. 
CFD grids with two densities, i.e., a coarse and a fine grid, were used. 
Each CFD cell contains around 2 and 25 particles for the fine and coarse grid cases, respectively. 
2.5mm diameter glass beads were used in these experiments. 
The particle properties \citep{link2005} and numerical settings for the simulations are listed in Table~\ref{tab:4}. 
In these simulations, the drag force model proposed in \citet{gidaspow1994} is employed, which applies the \citet{wen1966} and \citet{ergun1952} drag correlations for the dilute ($\varepsilon_f > 0.8$) and dense ($\varepsilon_f < 0.8$) regions, respectively. 
The \citet{ergun1952} and \citet{wen1966} models are used as they have been widely used to simulate the fluidization process \citep{zhu2007}.

\begin{table}[ht!]
    \caption{\label{tab:4}Particle properties and numerical settings in the simulation Case 2}
    \centering
    \begin{tabular}{c|c|c}
        \hline
        Property & Value & Unit \\
        \hline
        \hline
        Number of particles ($N_p$) & 24500 & - \\
        Density ($\rho_p$) & 2505 & kg/m3 \\
        Diameter ($D$) & 2.5 & mm \\
        Coefficient of normal restitution ($e$) & 0.97 & - \\
        Coulomb friction coefficient ($\mu$) & 0.1 & - \\
        Coarse CFD grid & $2\times30\times80$ & - \\
        Fine CFD grid & $6\times60\times300$ & - \\
        Inner inlet velocity ($U_{sp}$) & 30.0 & m/s \\
        Outer inlet velocity ($U_{bg}$) & 1.5 & m/s \\
        Cuboid-to-particle size ratio ($\theta_1$) & $1 \sim 6$ & - \\
        Point cloud density  ($\theta_2$) & $1 \sim 5$ & - \\
        Time step of CFD solver ($\Delta t_1$) & $5\times10^{-5}$ & s \\
        Time step of DEM solver ($\Delta t_2$) & $5\times10^{-7}$ & s \\
        Coupling interval ($M$) & $100 \Delta t_1$ & s \\
        \hline
    \end{tabular}
\end{table}

\begin{figure}[ht]
    \centering
    \begin{subfigure}[b]{0.49\textwidth}
        \centering
        \includegraphics[width=7.5cm]{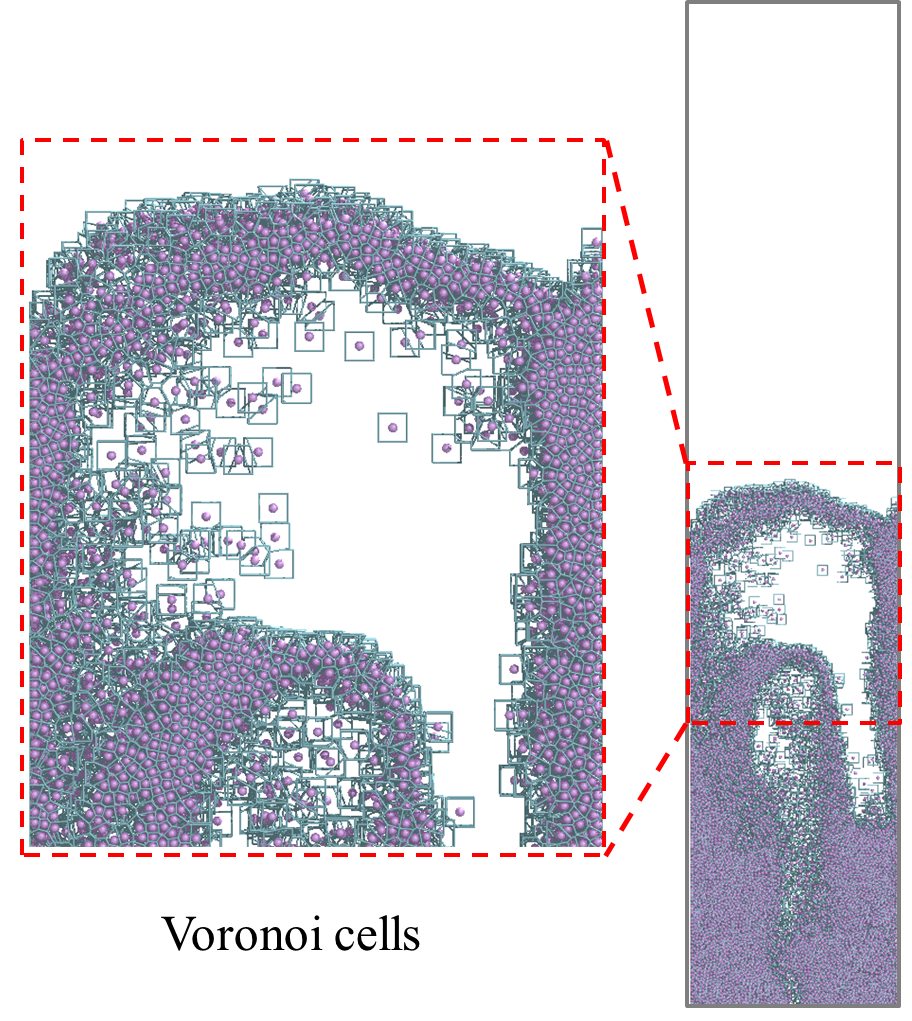}
        \caption{\label{fig:16a}The real particle distribution ($\theta_1 = 3.0$)}
    \end{subfigure}
    \hfill
    \begin{subfigure}[b]{0.49\textwidth}
        \centering
        \includegraphics[width=7.5cm]{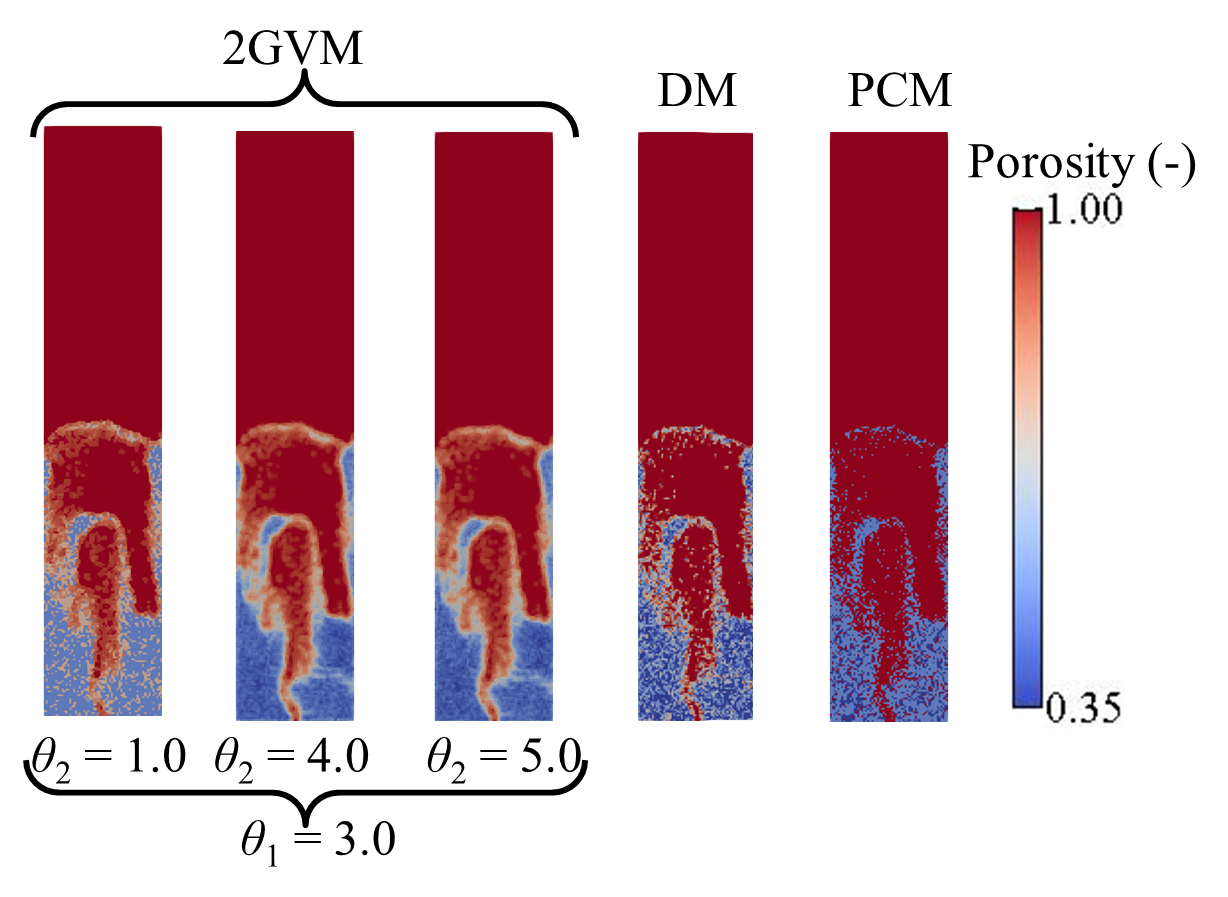}
        \caption{\label{fig:16b}Porosity distributions for fine CFD gird}
    \end{subfigure}
    \vfill
    \begin{subfigure}[b]{0.49\textwidth}
        \centering
        \includegraphics[width=7.5cm]{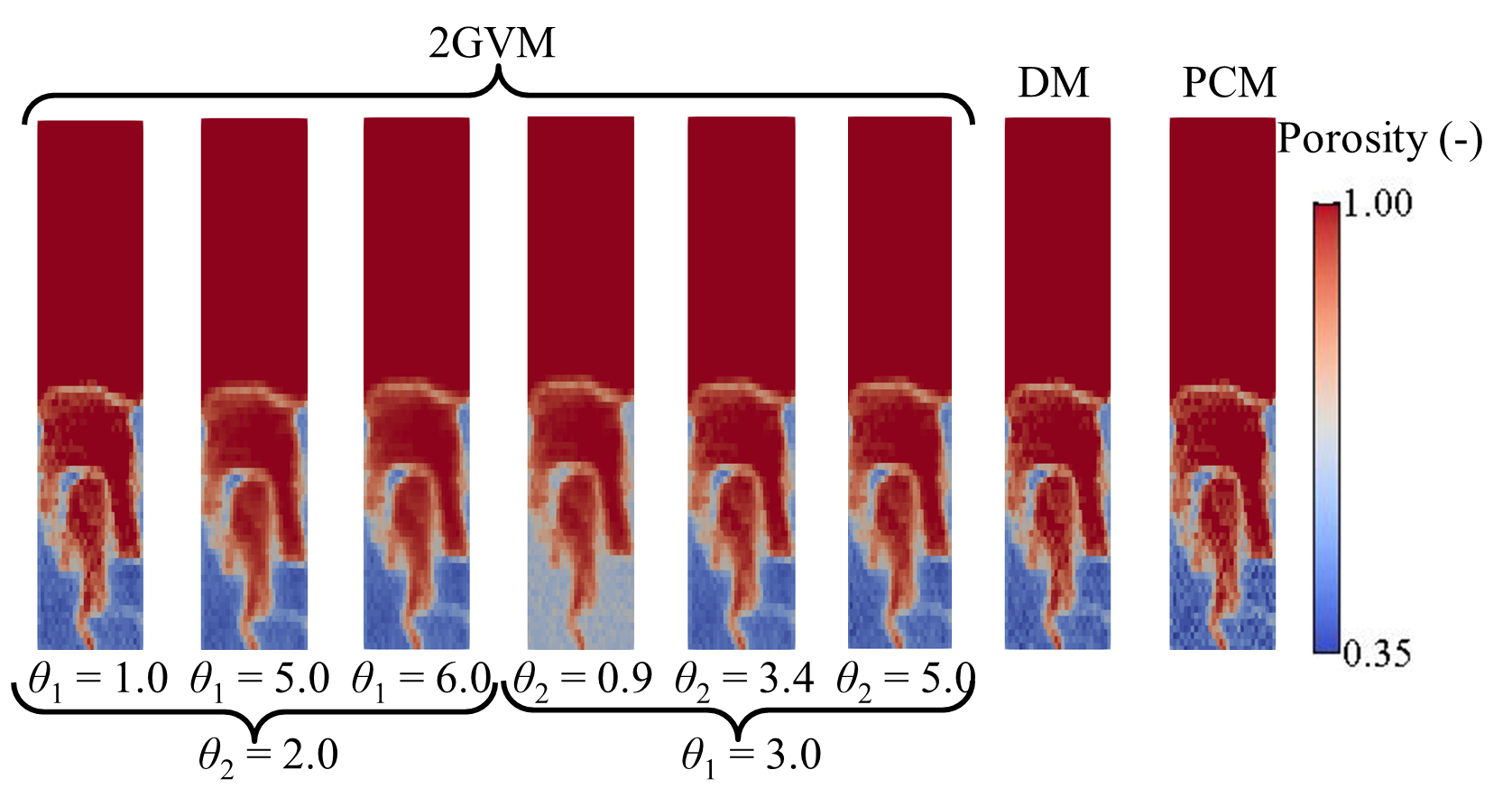}
        \caption{\label{fig:16c}Porosity distributions for coarse CFD gird}
    \end{subfigure}
    \hfill
    \begin{subfigure}[b]{0.49\textwidth}
        \centering
        \includegraphics[width=7.5cm]{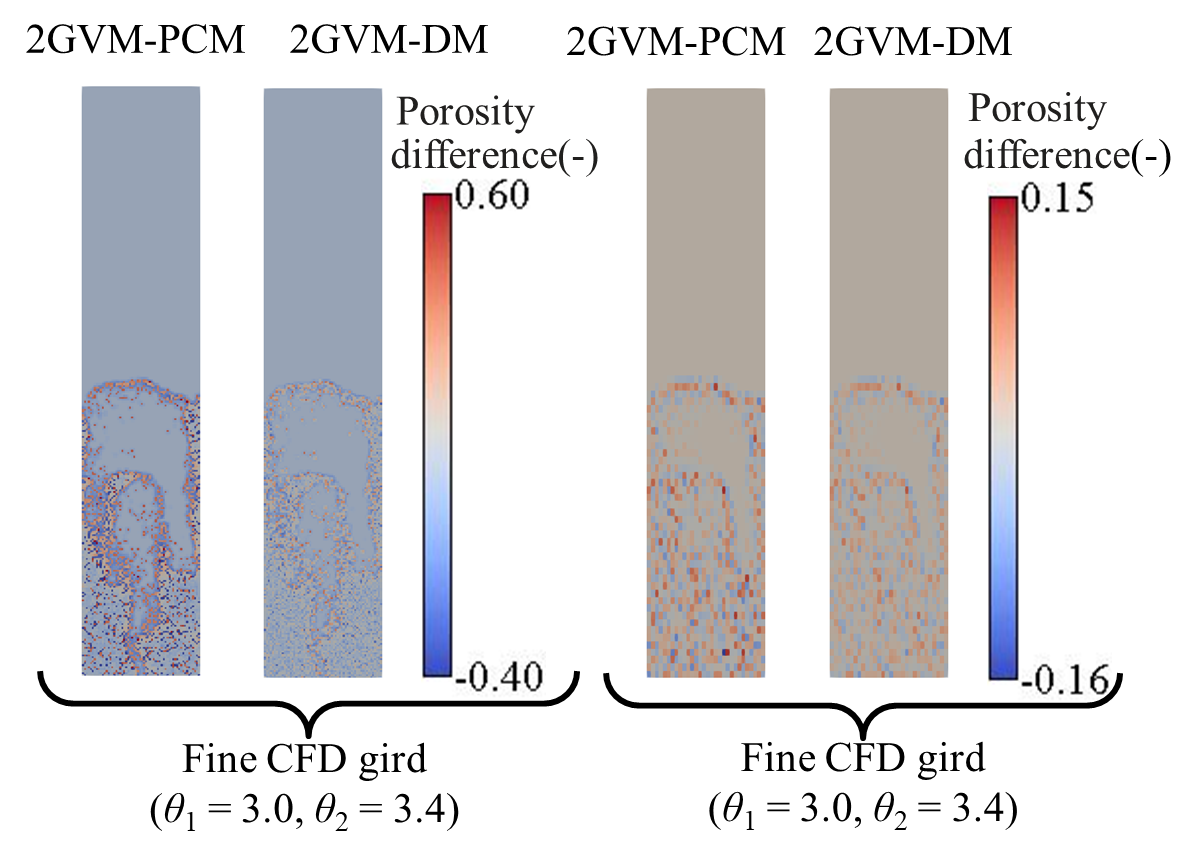}
        \caption{\label{fig:16d}Difference between porosity values calculated using 2GVM and those obtained using PCM \& DM }
    \end{subfigure}
    \caption{\label{fig:16}Plots of the same solids distribution with different coarse-graining methods. }
\end{figure}

In contrast to validation Case 1, there are regions of very low packing density in this system and so a bounding cuboid was applied to each particle and the sides of the Voronoi cells were not allowed to extend beyond the cuboid boundaries, as illustrated for a representative particle distribution in Fig.~\ref{fig:16a}. 

\begin{figure}[ht!]
    \centering
    \includegraphics[width=7cm]{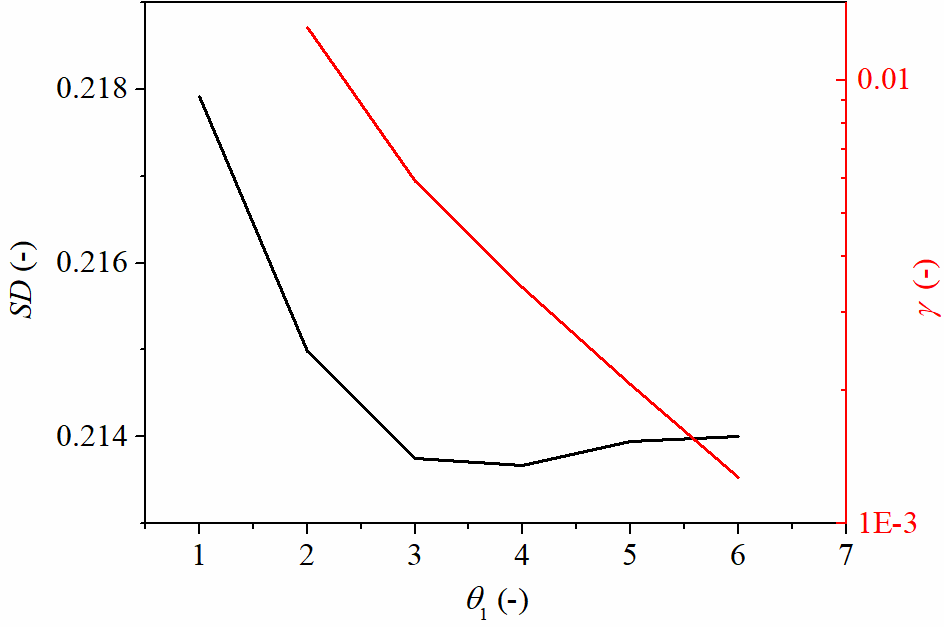}
    \caption{\label{fig:17}Variation of $SD$ and $\gamma$ with $\theta_1$ ($\theta_2 = 3.0$) for the coarse CFD grid}
\end{figure}

\begin{figure}[ht!]
    \centering
    \begin{subfigure}[b]{0.32\textwidth}
        \centering
        \includegraphics[width=5cm]{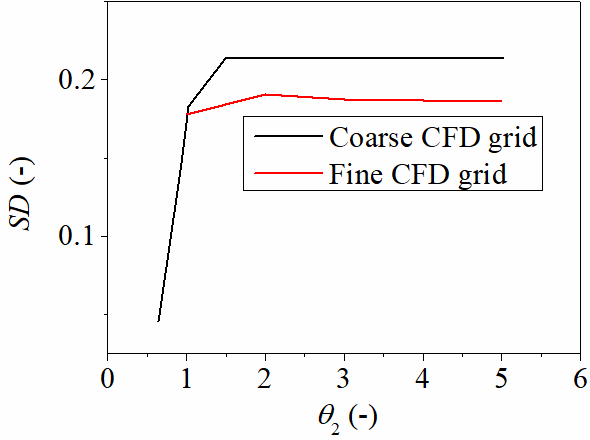}
        \caption{\label{fig:18a}Standard deviation}
    \end{subfigure}
    \hfill
    \begin{subfigure}[b]{0.32\textwidth}
        \centering
        \includegraphics[width=5cm]{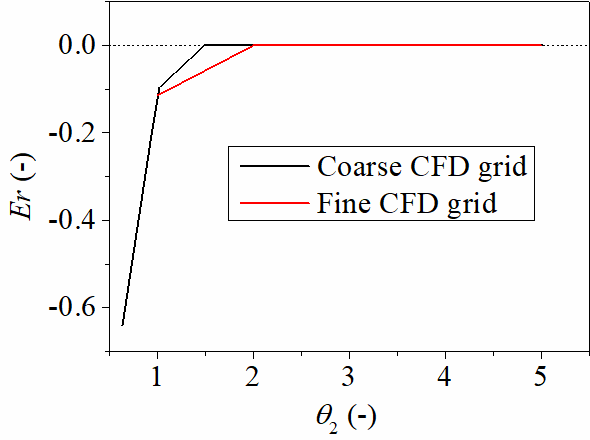}
        \caption{\label{fig:18b}Relative error}
    \end{subfigure}
    \hfill
    \begin{subfigure}[b]{0.32\textwidth}
        \centering
        \includegraphics[width=5cm]{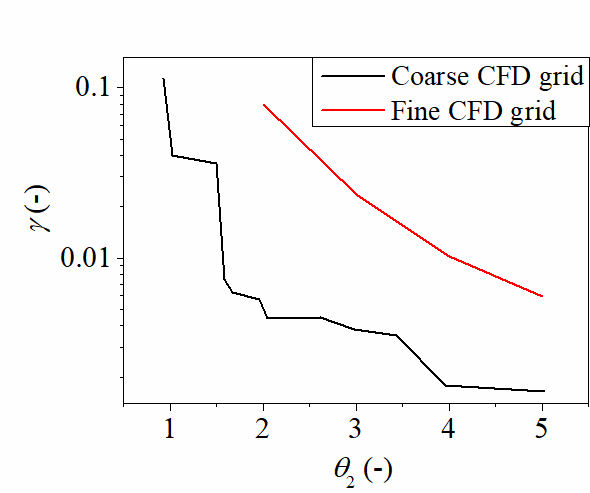}
        \caption{\label{fig:18c}Residual image error}
    \end{subfigure}
    \caption{\label{fig:18}Variation of $SD$, $Er$ and $\gamma$ with $\theta_2$  ($\theta_1 = 3.0$).}
\end{figure}

The effectiveness of different coarse-graining approaches in capturing the spatial distribution of porosity is shown in Fig.~\ref{fig:16b}-\ref{fig:16c} for the fine and coarse grid configurations. 
2GVM produces a substantially more smoother porosity field, which is most evident when considering the fine grid in Fig.~\ref{fig:16b}. 
Quantitative assessment of the differences between 2GVM, DM and PCM are presented in Fig.~\ref{fig:17} and Fig.~\ref{fig:18}. 
Fig.~\ref{fig:17} shows the variation in $SD$ and $\gamma$ for increasing $\theta_1$, while Fig. 18 shows the variation in $SD$, $Er$ and $\gamma$ with increasing $\theta_2$. 
This analysis indicates that the spatial distribution of porosity is insensitive to changes for $\theta_1 > 3.0$ and $\theta_2 > 3.0$, which is in alignment with the observations for validation Case 1. 
By comparing Fig.~\ref{fig:7a} with Fig.~\ref{fig:18a}, it is clear that the two figures indicate different initial trends in the $SD$ variation with $\theta_2$. 
These differences arise because when $\theta_2$ is small ($\theta_2\lesssim
1.5$), the point cloud density is low, therefore not all of the Voronoi cells may be detected by the point cloud. 
The resulting porosity field does not then correctly represent the true solid distribution, and the $Er$ in the total particle volume is high (see Fig.~\ref{fig:18b}). 
In these scenarios, the $SD$ of the porosity field can be either higher (Fig.~\ref{fig:7a}) or lower (Fig.~\ref{fig:18a}) than the final values ($\theta_2>3.0$), resulting in different trends in the $SD$ curves.  
These data confirm that the method is not effective if a $\theta_2$ value less than 1.5 is selected.

\begin{figure}[ht!]
    \centering
    \textbf{Fine CFD grid:}
    \vfill
    \begin{subfigure}[b]{0.32\textwidth}
        \centering
        \includegraphics[width=5cm]{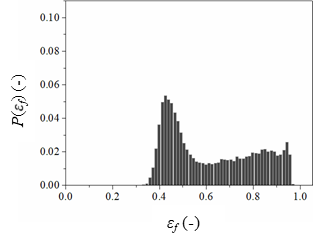}
        \caption{\label{fig:19a}2GVM ($\theta_1 = 3.0$, $\theta_2 = 2.0$)}
    \end{subfigure}
    \hfill
    \begin{subfigure}[b]{0.32\textwidth}
        \centering
        \includegraphics[width=5cm]{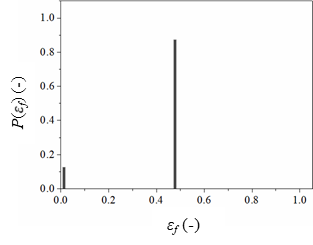}
        \caption{\label{fig:19b}PCM}
    \end{subfigure}
    \hfill
    \begin{subfigure}[b]{0.32\textwidth}
        \centering
        \includegraphics[width=5cm]{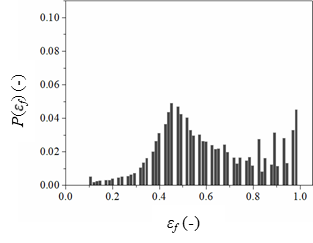}
        \caption{\label{fig:19c}DM}
    \end{subfigure}
    \vfill

    \textbf{Coarse CFD grid:}
    \vfill
    \centering
    \begin{subfigure}[b]{0.32\textwidth}
        \centering
        \includegraphics[width=5cm]{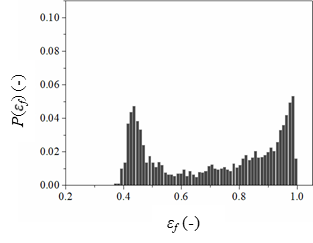}
        \caption{\label{fig:19d}2GVM ($\theta_1 = 3.0$, $\theta_2 = 2.0$)}
    \end{subfigure}
    \hfill
    \begin{subfigure}[b]{0.32\textwidth}
        \centering
        \includegraphics[width=5cm]{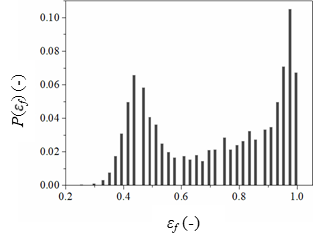}
        \caption{\label{fig:19e}PCM}
    \end{subfigure}
    \hfill
    \begin{subfigure}[b]{0.32\textwidth}
        \centering
        \includegraphics[width=5cm]{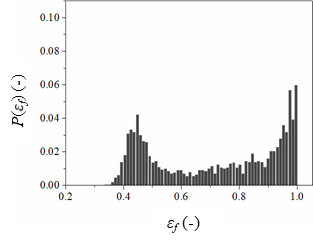}
        \caption{\label{fig:19f}DM}
    \end{subfigure}
    \caption{\label{fig:19}Histogram distributions of the porosity values in CFD grid from different coarse-graining methods.}
\end{figure}

\begin{figure}[ht!]
    \centering
    \includegraphics[width=7cm]{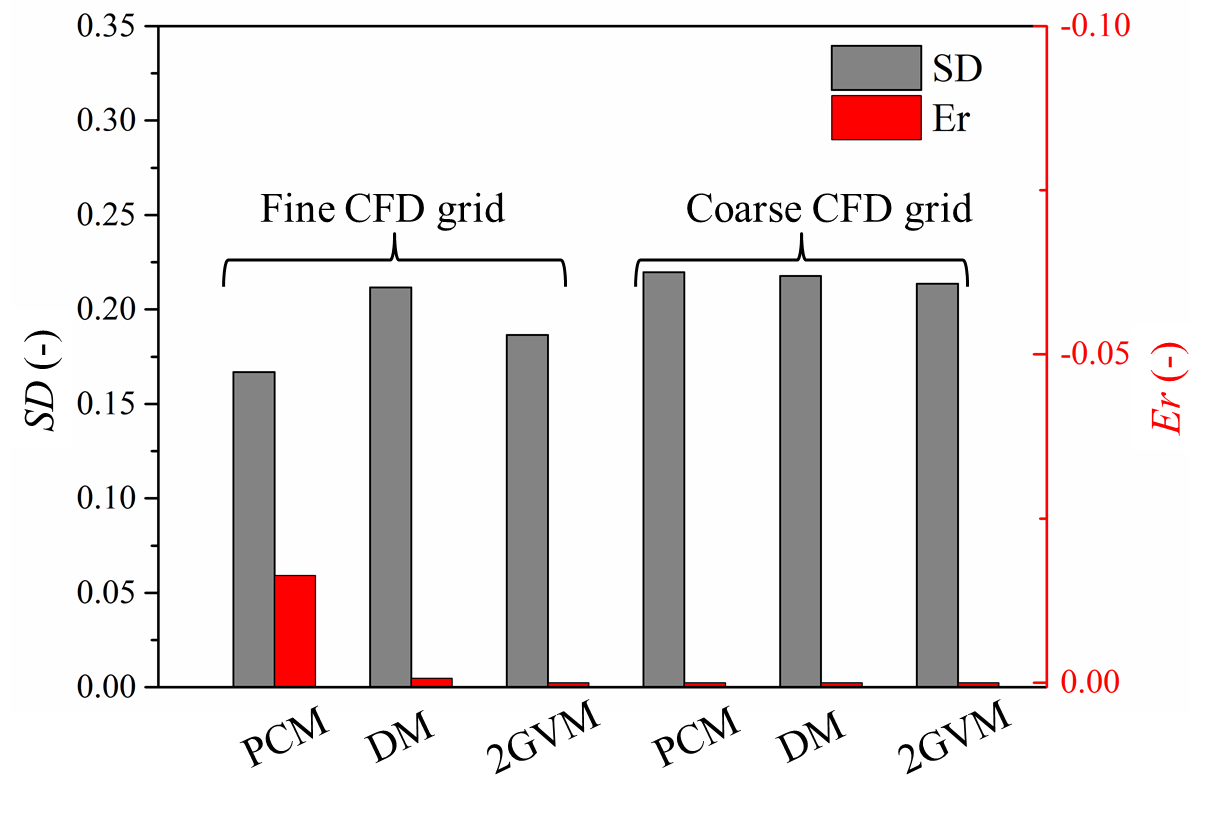}
    \caption{\label{fig:20}Comparison of the $SD$ of the porosity distributions from different coarse-graining methods}
\end{figure}

In order to compare the porosity fields calculated from 2GVM, PCM and DM, Fig.~\ref{fig:19} shows the histogram of the porosity from these methods, while the $SD$ and $Er$ of the porosity values are shown in Fig.~\ref{fig:20}. 
Only the CFD cells occupied by particles were included in the statistics. 
From Fig.~\ref{fig:19}, the distributions of the porosity predicted from PCM and DM are not continuous for the fine CFD grid case, as indicated by the relatively large jumps that exist between cells. 
In addition, particle volume is not conserved when PCM is used with a fine grid (see Fig.~\ref{fig:20}). 
The nonuniformity and jumps in the porosity field from PCM and DM can be confirmed by the visual inspection of Fig.~\ref{fig:16}. 
A coarse CFD grid produces more similar results between the various methods. As the CFD grid size is increased, the three methods provide SD values for the porosity field that are very close to each other and the total solid volume is now effectively conserved in all cases ($Er \approx 0$). 
Owing to the low cell-to-particle size ratio in the fine grid case, the accuracy of porosity field predicted by PCM and DM is low. 
By a similar logic, the differences in the porosity distributions estimated from the three coarse-graining methods are not expected to be pronounced for the coarse grid case. 
The above trends were also observed in validation Case 1. 

Fig.~\ref{fig:16d} shows the porosity differences, which are obtained by subtracting the porosity values in each CFD cell obtained using the PCM and DM from the 2GVM values, for the fine and coarse CFD grid cases. 
From Fig.~\ref{fig:16d}, the porosity differences in the fine CFD grid case are evident in the dilute particle region between the solid and fluid phase. 
For the coarse CFD grid case, the magnitude of the image difference is obviously lower and is uniformly distributed in the region that occupied by the particles. 
As mentioned above, PCM and DM have limitations when calculating porosity for the fine CFD grid but are suitable for the coarse CFD grid cases, 2GVM is suitable for both fine and coarse CFD grid cases. 
Accordingly, Fig.~\ref{fig:16d} indicates that 2GVM mainly improves the porosity field in the dilute flow regions of fine CFD grid case.

\clearpage

\begin{figure}[ht!]
    \centering
    \includegraphics[width=7cm]{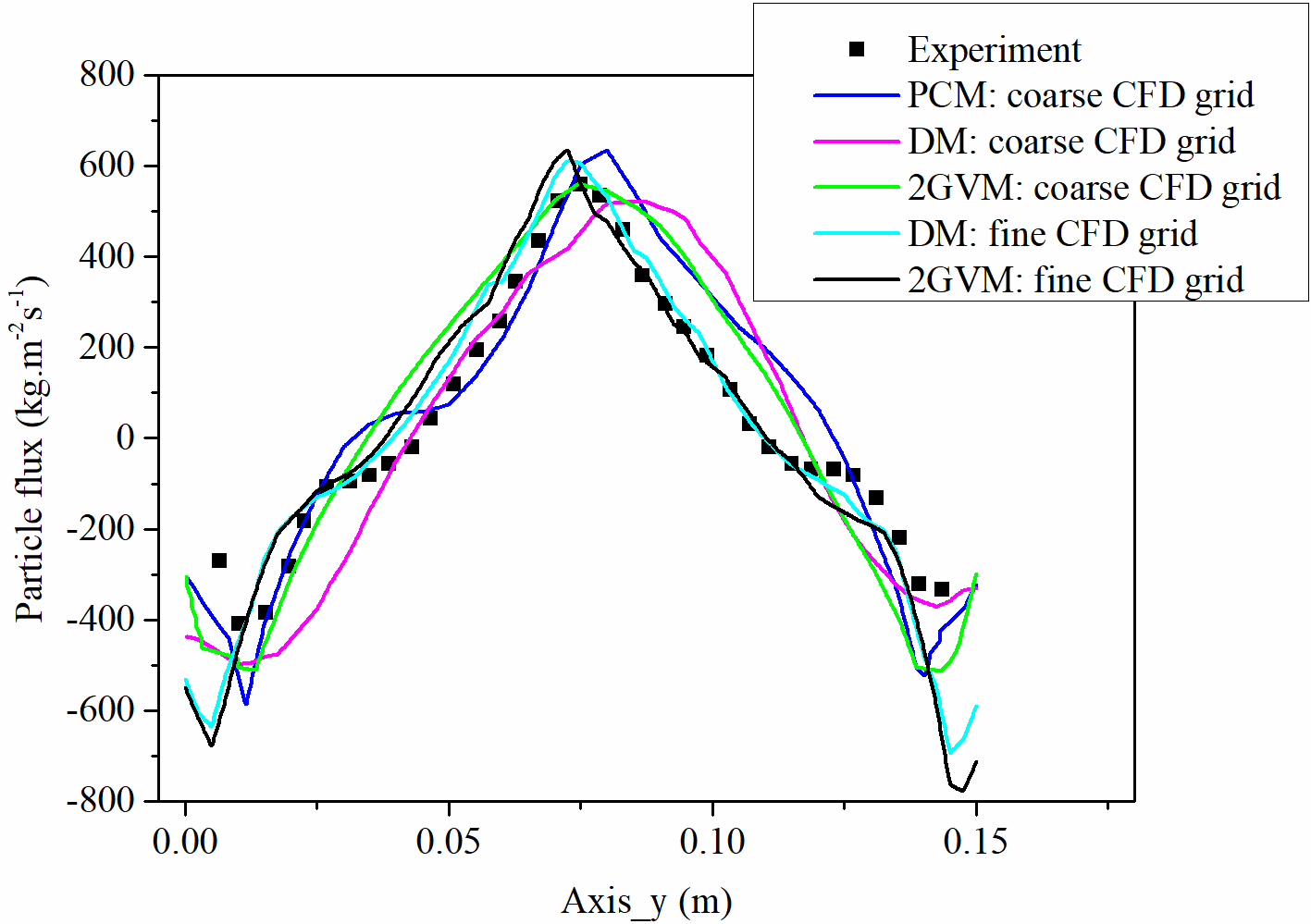}
    \caption{\label{fig:21}Comparison of the particle flux profiles from the experiment and simulations}
\end{figure}

Fig.~\ref{fig:21} compares the particle flux profiles obtained from the experimental image analysis of the CCD camera and the CFD-DEM simulations. 
The particle flux is the averaged value over 5 seconds of physical time in the simulations. 
A good agreement is observed for all of the simulation cases. 
In validation Case 1, we were able to demonstrate that 2GVM both provides a smooth porosity field as well as a correct estimation of the individual particle drag forces at the particle-scale. 
For this validation case, only the particle flow information at the macro-level (such as the particle flux) are extracted from experimental measurements and available for the comparison with numerical simulation. 
The advantage of 2GVM is not obvious in such conditions as the conventional PCM and DM are able to give a reliable prediction of the macro-level flow dynamics. 
The objective of the current study was to validate/benchmark the methods using data from a well-constrained problem for which reliable experimental data are available. 
Future studies will consider application of the 2GVM model to polydisperse systems involving fluidised beds. 

\section{\label{sec:5}Conclusions}

This study introduced a new coarse-grained approach for CFD-DEM simulations (2GVM). 
The new approach used a radical Voronoi tessellation so that the fluid-particle interaction force can be calculated using local porosity data. 
A point cloud is used to map field variables between the CFD and DEM solvers. 
Two different, but typical validation cases were used to evaluate the performance of the proposed method. 

The following main conclusions are drawn based on the two verification studies considered:
\begin{enumerate}
    \item The new coarse-graining method (2GVM) conserves data, is grid-independent and can generate smooth porosity field in the CFD side without using any smoothing or thresholding method.
    \item 2GVM provides a more reasonable estimation of the individual particle porosity for polydisperse systems, and results in an accurate prediction on the individual particle drag when compared with data obtained using a fully resolved IBM simulations.
    \item The approach taken to modify the Voronoi tessellation with an outer bounding cuboid can be easily applied to the applications of the fluidised bed, where the solid concentration is non-uniformly distributed.
\end{enumerate}
This new coarse-graining approach is particularly relevant to cases where the variation in the fluid-particle interaction force with particle size is important, for example in the simulation of internal instability in the embankment soils or any application where particle segregation occurs.

\section*{Acknowledgements}

This work was financially supported by EPSRC grant  EP/P010393/1. Key development work by Dr. Edward Smith was funded by the embedded CSE programme of the ARCHER UK National Supercomputing Service (http://www.archer.ac.uk). 

\bibliographystyle{ms}
\bibliography{ms}

\end{document}